\documentclass[%
 reprint,
superscriptaddress,
 amsmath,amssymb,
 aps,
]{revtex4-1}

\usepackage{color}
\usepackage{graphicx}
\usepackage{dcolumn}
\usepackage{bm}
\usepackage[T1]{fontenc}

\begin{document}

\preprint{APS/123-QED}

\title{The Deconfinement Phase Transition in Proto-Neutron-Star Matter}

\author{J. Roark}
\affiliation{Department of Physics, Kent State University, Kent, OH 44243 USA}

\author{V. Dexheimer}
\affiliation{Department of Physics, Kent State University, Kent, OH 44243 USA}



\date{\today}

\begin{abstract}
In this work, we study in detail the deconfinement phase transition that takes place in hot/dense nuclear matter in the context of neutron stars and proto-neutron stars (in which lepton fraction is fixed). The possibility of different mixtures of phases with different locally and globally conserved quantities is considered in each case. For this purpose, the Chiral Mean Field (CMF) model, an effective relativistic model that includes self-consistent chiral symmetry restoration and deconfinement to quark matter, is employed. Finally, we compare our results with blue results provided by PQCD for different temperatures and conditions.
\end{abstract}                       
                              
\maketitle


\section{Introduction}

The core of a neutron star can roughly be described as a sea of infinite nuclear matter -- protons and neutrons strongly interacting at low temperature (relatively speaking, in the MeV scale) and high density. As we dive deeper towards the inner core, hyperons should appear and, eventually, hadrons become packed so tightly that they can ``dissolve'' and quark deconfinement occurs (i.e., the baryons no longer act like clusters of quarks and instead behave like a sea of disassociated quarks). Thus, neutron-star matter can either exist in the hadronic phase, the quark phase, or in a state consisting of a mixture of these two phases, in which quantities of interest can be conserved globally amongst the two phases instead of simply locally within each phase (see Refs.~\cite{Lukacs:1986hu, Heinz:1987sj, Glendenning:1992vb} and Ref.~\cite{Hempel:2013tfa} with references therein for details).

Throughout their lives, a large part of the cooling experienced by neutron stars takes the form of neutrino emission. But, early on, just after its progenitor supernova event, the hot, dense medium of young neutron stars (or "proto-neutron stars") causes the mean free path of the neutrinos to drop dramatically (i.e., less than the radius of the star) \cite{Prakash:1996xs, Reddy:1997yr, Shen:2003ih, Pastore:2014yua}. Thus, the primary difference between the description of neutron-star (NS) matter and proto-neuron-star (PNS) matter lies in fact that, in the latter case, lepton fraction must be fixed. In order to ensure stability in each system (i.e., to keep matter gravitationally bound), it is also necessary for our description to conserve electric charge (more specifically, to keep the system electrically neutral) in both neutron and proto-neutron stars. Ref$.$ \cite{Carter:2001ji} presents in Fig$.$ 1 a sketch of the path of neutron stars throughout their temporal evolution in the QCD phase diagram. For a review of the thermal properties of bulk hadronic matter, see for example Refs$.$ \cite{Constantinou:2014hha,Du:2018vyp}. For dynamical simulations including hadron-quark phase transitions, see for example the recent Refs.~\cite{Nakazato:2008su, Sagert:2008ka, Nakazato:2010ue, Fischer:2010wp, 2012MNRAS.423.1652O, Fischer:2017lag}, for studies of phase transitions in proto-neutron star matter, see for example Refs.~\cite{Drago:1997tn, Lugones:2011xv, Bombaci:2011mx, Grunfeld:2018lxb}, and for studies on quark-pasta structures, see for example Refs.~\cite{Yasutake:2014oxa, Ayriyan:2017tvl}.

For the purposes of this paper, electric charge (Q) and lepton fraction ($Y_l$) will serve as our conserved quantities of interest (in addition to baryon number conservation). These quantities can either be strictly conserved locally within each phase (which leads to a "congruent" phase transition, where there is no phase coexistence and the phases are distinctly separated) or globally amongst a mixture of phases (which leads to a "non-congruent" phase transition, where there is a phase coexistence of two or more macroscopic phases with different chemical compositions) \cite{Reid:1966, Landau:1969, Greiner:1987tg, Clarke:1994, Iosilevskiy:2003, Iosilevskiy:2010qr}. Therefore, the following scenarios are considered: neutron-star matter in the case of locally conserved electric charge (NS LCN), neutron-star matter in the case of globally conserved electric charge (NS GCN), proto-neutron-star matter in the case of locally conserved electric charge and lepton fraction (PNS LCN L$Y_l$), proto-neutron-star matter in the case of locally conserved electric charge and globally conserved lepton fraction (PNS LCN G$Y_l$), and proto-neutron-star matter in the case globally conserved electric charge \textit{and} lepton fraction (PNS GCN G$Y_l$). We extended our NS and PNS calculations to high temperatures for the sake of comparing them with each other and comparing with the charge fraction constrained matter from Ref$.$ \cite{Hempel:2013tfa}.

Neutron-star matter exists in the low temperature but relatively high density regime and, because of this, common methods that describe high-energy matter cannot be directly applied. For instance, perturbative QCD (PQCD) is applicable to systems involving weaker interactions than those present inside most neutron and proto-neutron stars \cite{Andersen:2002jz, Fraga:2013qra, Kurkela:2016was}. On the other hand, lattice QCD exhibits the sign problem that arises at nonzero baryon density \cite{Muroya:2003qs, deForcrand:2010ys}. Therefore, we choose to employ an effective model for our description, namely, the Chiral Mean Field (CMF) model, which can describe properties of hot/dense nuclear matter, such as chiral symmetry restoration and deconfinement to quark matter. Nevertheless, we calibrate the CMF model to agree with lattice QCD results and draw comparisons with PQCD results in the relevant limits. In the past, we have addressed the influence of lepton fraction in the purely hadronic version of the CMF model \cite{Dexheimer:2008ax,Franzon:2016iai}. On the other hand, we have also studied in detail phase diagrams built within the CMF model under the conditions of charge neutrality and chemical equilibrium (for neutron stars), as well as charge fraction without charge neutrality and without leptons (to study heavy-ion collisions) \cite{Dexheimer:2009hi,Hempel:2013tfa,Dexheimer:2012qk,Dexheimer:2017ecc}.

\section{Formalism}

The CMF model is based on a non-linear realization of the SU(3) chiral sigma model. It is a relativistic model constructed from symmetry relations, which allows it to be chirally invariant in the expected regime \cite{Papazoglou:1998vr}. The baryon and quark masses are generated by interactions with the medium and, therefore, decrease with temperature and/or density. The Lagrangian density of the CMF model in the mean field approximation reads \cite{Dexheimer:2008ax, Dexheimer:2009hi}:
\begin{eqnarray}
&L = L_{Kin} + L_{Int} + L_{Self} + L_{SB} - U,&
\end{eqnarray}
where, besides the kinetic energy term for hadrons, quarks, and leptons (included to ensure charge neutrality) the terms remaining are:
\begin{eqnarray}
&L_{Int} = -\sum_i \bar{\psi_i} [\gamma_0 (g_{i \omega} \omega + g_{i \phi} \phi + g_{i \rho} \tau_3 \rho) + M_i^*] \psi_i,& \nonumber\\ \\
&L_{Self} = \frac{1}{2} (m_\omega^2 \omega^2 + m_\rho^2 \rho^2 + m_\phi^2 \phi^2)& \nonumber \\
&+ g_4 \left(\omega^4 + \frac{\phi^4}{4} + 3 \omega^2 \phi^2 + \frac{4 \omega^3 \phi}{\sqrt{2}} + \frac{2 \omega \phi^3}{\sqrt{2}}\right)& \nonumber \\
&- k_0 (\sigma^2 + \zeta^2 + \delta^2) - k_1 (\sigma^2 + \zeta^2 + \delta^2)^2& \nonumber \\
&- k_2 \left(\frac{\sigma^4}{2}+\frac{\delta^4}{2} + 3 \sigma^2 \delta^2 + \zeta^4\right) - k_3 (\sigma^2 - \delta^2) \zeta& \nonumber \\
&- k_4 \ \ln{\frac{(\sigma^2 - \delta^2) \zeta}{\sigma_0^2 \zeta_0}},& \\ \nonumber \\
&L_{SB} = -m_\pi^2 f_\pi \sigma - \left(\sqrt{2} m_k^ 2f_k - \frac{1}{\sqrt{2}} m_\pi^ 2 f_\pi\right) \zeta,& \\ \nonumber \\
&U = (a_o T^4 + a_1 \mu_B^4 + a_2 T^2 \mu_B^2) \Phi^2& \nonumber \\
&+ a_3 T_o^4 \ \ln{(1 - 6 \Phi^2 + 8 \Phi^3 -3 \Phi^4)}.&
\end{eqnarray}
Respectively, these represent the interactions between baryons (and quarks) and vector/scalar mesons, the self interactions of scalar and vector mesons, an explicit chiral symmetry breaking term responsible for producing the masses of the pseudo-scalar mesons, and the effective potential for the scalar field $\Phi$, an analogy to the Polyakov loop in the PNJL approach \cite{Ratti:2006ka, Roessner:2006xn}. The underlying flavor symmetry of the model is SU(3) and the index $i$ denotes the baryon octet, the three light quarks, electrons, muons, and electron neutrinos. The mesons included are the vector-isoscalars $\omega$ and $\phi$ (strange quark-antiquark state), the vector-isovector $\rho$, the scalar-isoscalars $\sigma$ and $\zeta$ (strange quark-antiquark state), and the scalar-isovector $\delta$. The isovector mesons affect isospin-asymmetric matter and are, consequently, important for neutron star physics.

The coupling constants of the hadronic part of the model are shown in Table~\ref{constants1}. They were fitted to reproduce the vacuum masses of baryons and mesons, nuclear saturation properties (density $\rho_0=0.15$ fm$^{-3}$, binding energy per nucleon $B/A=-16$ MeV, compressibility $K=300$ MeV), the asymmetry energy ($E_{sym}=30$ MeV) and its slope ($L=88$ MeV), and reasonable values for the hyperon potentials ($U_\Lambda=-28.00$ MeV, $U_\Sigma=5$ MeV, $U_\Xi=-18$ MeV). The reproduced critical point for the nuclear liquid-gas phase transition lies at $T_c=16.4$ MeV, $\mu_{B,c}=910$ MeV. The vacuum expectation values of the scalar mesons are constrained by reproducing the pion and kaon decay constants. It should be noted that all coupling constants for the leptons are zero.

\begin{table}
	\caption{\label{constants1}
		Coupling constants for the CMF model containing only baryons ($\chi_0=401.93$ MeV).}
	\begin{ruledtabular}
		\begin{tabular}{ccc}
			$g_{N \omega}=11.90$ & $g_{N \phi}=0$ & $g_{N \rho}=4.03$ \\
			$g_{N \sigma}=-9.83$  & $g_{N \zeta}=1.22$ & $g_{N \delta}=-2.34$ \\
			$g_{\Lambda \omega}=7.93$ & $g_{\Lambda \phi}=-7.32$ & $g_{\Lambda \rho}=0$ \\
			$g_{\Lambda \sigma}=-5.52$ & $g_{\Lambda \zeta}=-2.30$ & $g_{\Lambda \delta}=0$ \\
			$k_0=1.19 \chi_0^2$ & $k_1=-1.40$ & $k_2=5.55$ \\
			$k_3=2.65 \chi_0$ & $k_4=-0.02 \chi_0^4$ & $g_4=38.90$ \\
		\end{tabular}
	\end{ruledtabular}
\end{table}

\begin{table}
	\caption{\label{constants2}
		Additional coupling constants for quark section of the model.}
	\begin{ruledtabular}
		\begin{tabular}{ccc}
			$g_{u \omega}=0$ & $g_{u \phi}=0$ & $g_{u \rho}=0$ \\
			$g_{u \sigma}=-3$ & $g_{u \zeta}=0$ & $g_{u \delta}=0$ \\
			$g_{d \omega}=0$ & $g_{d \phi}=0$ & $g_{d \rho}=0$ \\
			$g_{d \sigma}=-3$ & $g_{d \zeta}=0$ & $g_{d \delta}=0$ \\
			$g_{s \omega}=0$ & $g_{s \phi}=0$ & $g_{s \rho}=0$ \\
			$g_{s \sigma}=0$ & $g_{s \zeta}=-3$ & $g_{s \delta}=0$ \\
			$a_0=-1.85$ & $a_1=-1.44$x$10^{-3}$ & $a_2=-0.08$ \\
			$a_3=-0.40$ & $g_{B\Phi}=1500$ MeV & $g_{q\Phi}=500$ MeV \\
			$T_0=200$ MeV & $T_0$ (gauge) $=270$ MeV & \\
		\end{tabular}
	\end{ruledtabular}
\end{table}

The mesons are treated as classical fields within the mean-field approximation. Finite-temperature calculations include the heat bath of hadronic and quark quasiparticles within the grand canonical ensemble. The grand potential of the system is defined as:
\begin{eqnarray}
&\frac{\Omega}{V} = -L_{Int} - L_{Self} - L_{SB} - L_{Vac} + U &\nonumber\\& + T \sum_i \frac{\gamma_i}{(2 \pi)^3} \int_{0}^{\infty} \, d^3k \, \ln(1 + e^{-\frac{1}{T}(E_i^*(k) \mp \mu_i^*)}),&
\end{eqnarray}
where $L_{Vac}$ is the vacuum energy, $\gamma_i$ is the fermionic degeneracy, $E_{i}^* (k) = \sqrt{k^2 + {M^*_i}^2}$ is the single particle effective energy, $\mu_i^* = \mu_i - g_{i\omega} \omega - g_{\phi} \phi - g_{i\rho} \tau_3 \rho$ is the effective chemical potential of each species, and the $\mp$ in the exponential function refers to particles and antiparticles, respectively. The chemical potential for each species $\mu_i$ is determined by the chemical equilibrium conditions.

Due to their interactions with the mean field of mesons and the field $\Phi$, the effective masses of baryons and quarks take the form:
\begin{eqnarray}
&M_{B}^* = g_{B \sigma} \sigma + g_{B \delta} \tau_3 \delta + g_{B \zeta} \zeta + M_{0_B} + g_{B \Phi} \Phi^2,& \\ \nonumber \\
&M_{q}^* = g_{q \sigma} \sigma + g_{q \delta} \tau_3 \delta + g_{q \zeta} \zeta + M_{0_q} + g_{q \Phi}(1 - \Phi),&
\end{eqnarray}
where the bare masses are $M_0=150$ MeV for nucleons, $354.91$ MeV for hyperons, $5$ MeV for up and down quarks, and $150$ MeV for strange quarks (see Table~\ref{constants2} for more coupling constants for the quark sector of the model). Notice that for small values of $\Phi$, $M_{B}^*$ is small while $M_{q}^*$ is very large. This essentially indicates that, for small $\Phi$ values, the presence of baryons is promoted while quarks are suppressed, and vice versa. In this sense, $\Phi$ acts as an order parameter for deconfinement (or, in the case of a mixture of phases, as an indicator as to which phase is dominant).

The coupling constants for the quark sector of the model are chosen to reproduce lattice data as well as known information about the phase diagram. The lattice data includes a first order phase transition at $T=270$ MeV and a pressure functional $P(T)$ similar to Refs$.$ \cite{Ratti:2005jh, Roessner:2006xn} at $\mu=0$ for pure gauge, a crossover at vanishing chemical potential with a transition temperature of $171$ MeV (determined as the peak of the change of the chiral condensate and $\Phi$), and the location of the critical end-point (at $\mu_c=354$ MeV, $T_c=167$ MeV for symmetric matter in accordance with one of the existent calculations \cite{Fodor:2004nz}). The phase diagram information includes a continuous first order phase transition line that terminates on the zero temperature axis at four times saturation density. The numerical code for the CMF model solves a set of equations for each baryon chemical potential and temperature. Those include an equation of motion for each meson. Additional constraints such as charge neutrality, fixed lepton fraction, and fixed entropy require additional equations.

It should be mentioned at this point that the CMF model allows for the existence of soluted quarks in the hadronic phase and soluted hadrons in the quark phase. This is true even in the case of congruent phase transitions with no mixture of phases. Regardless, quarks will always give the dominant contribution in the quark phase, and hadrons in the hadronic phase, due to the fact that the effective masses of both quarks and hadrons are a function of $\Phi$. We assume that this inter-penetration of quarks and hadrons is indeed physical, and is required to achieve the crossover transition at low $\mu_B$ values \cite{Aoki:2006we}.

For each fermionic species in the system, we define its chemical potential as
\begin{eqnarray}
\mu_i = Q_{B,i} \ \mu_B + Q_i \ (\mu_Q + \mu_l) +  Q_{l,i} \ \mu_l,
\end{eqnarray}
where $\mu_B$, $\mu_Q$, and $\mu_l$ represent the chemical potentials corresponding to the conserved quantities of baryon number, electric charge, and lepton fraction, respectively. The values $Q_{B,i}$, $Q_i$, and $Q_{l,i}$ are the baryon charge, electric charge, and lepton charge of a particular species $i$. Note that Eq.~(9) can be rewritten in a more intuitive way,
\begin{eqnarray}
\mu_i = Q_{B,i} \ \mu_B + Q_i \ \mu_Q' +  Q_{l,i} \ \mu_l,
\end{eqnarray}
by redefining the charged chemical potential as $\mu_Q' = \mu_Q + \mu_l$. The total electric charge density is calculated as
\begin{eqnarray}
\frac{Q}{V} = \sum_i Q_i n_i,
\end{eqnarray}
where $n_i$ is the number density of particle species $i$.

Note that, in the case that some quantity is being conserved locally, the value of that quantity is the same in both phases, by definition. But this is not true for the corresponding chemical potential (i.e., $\mu_{j,H} \neq \mu_{j,Q}$). On the other hand, in the case that a quantity is being conserved \textit{globally}, the value is different in each phase, by definition, but the corresponding chemical potentials are equal, defining an additional equilibrium condition (i.e., $\mu_{j,H} = \mu_{j,Q}$).

The lepton fraction $Y_l$ is defined as the number leptons in our system divided by the number of baryons:
\begin{eqnarray}
Y_l = \frac{L}{B} = \frac{\sum_i Q_{l,i} \ n_i}{\sum_i Q_{B,i} \ n_i} = \frac{n_l}{n_B^o}.
\end{eqnarray}
For our purposes, $n_l = n_e + n_\nu$ (the sum of the electron number density and the electron neutrino number density). Thus, when we are conserving lepton fraction, this conservation applies to neutrinos \textit{and} electrons. Note that $n_B^o = \sum_i Q_{B,i} n_i$ is not the same as the baryon number density $n_B$, as the latter comes from the derivative of the pressure with respect to the baryon chemical potential and, therefore, also contains a contribution from the potential $U$ for $\Phi$, namely, $n_\Phi$ (when quarks are present). For the purposes of this paper, when $Y_l$ is being fixed, its value is held at 0.4 \cite{Fischer:2009af,Huedepohl:2009wh}. This typical value comes from numerical simulations of proto-neutron-star evolution. A similar quantity, $Y_Q$ (the electric charge per baryon) is defined as
\begin{eqnarray}
Y_Q = \frac{Q}{B} = \frac{\sum_i Q_i \ n_i}{\sum_i Q_{B,i} \ n_i}.
\end{eqnarray}
As mentioned previously, this quantity must be set to zero to ensure electric charge neutrality, as a significant net excess of electric charges could not be kept in the star by gravity.

In order to take into account the presence of neutrinos in the appropriate scenarios, it is also advantageous to define a modified chemical potential
\begin{eqnarray}
\tilde{\mu} = \mu_B + Y_Q \ (\mu_Q + \mu_l) + Y_l \ \mu_l,
\end{eqnarray}
which is equal to the Gibbs free energy per baryon. This value comes from the definition of the energy density of the system,
\begin{eqnarray}
\varepsilon = -P + T s + \sum_i \mu_i n_i + \mu_B n_\Phi,
\end{eqnarray}
where, from Eqs.~(9-13), it can be shown that $\sum_i \mu_i n_i = [\mu_B + Y_Q \ (\mu_Q + \mu_l) + Y_l \ \mu_l] \left(\sum_i Q_{B,i} \ n_i\right)$. Because of the condition of electric charge neutrality, $Y_Q$ is zero in all cases studied in this work, causing Eq.~(14) to read $\tilde{\mu} = \mu_B + Y_l \ \mu_l$ and $\sum_i \mu_i n_i = \tilde{\mu} \left(\sum_i Q_{B,i} \ n_i\right)$. In the case that lepton fraction is \textit{not} fixed but the condition of electric charge neutrality is still enforced, $\mu_l=0$ (because leptons are free to leave the system), $\tilde{\mu}=\mu_B$ and $\sum_i \mu_i n_i = \mu_B \left(\sum_i Q_{B,i} \ n_i\right)$.

When studying scenarios involving mixtures of phases, it becomes important to define the continuous variable $\lambda$ (the volume fraction of quarks). When $\lambda=0$, the mixture of phases is entirely composed of hadronic matter and, when $\lambda=1$, it consists entirely of quark matter.

\begin{figure}[t!]
	\vspace{3mm}
	\includegraphics[trim={1.4cm 0 0 2.cm},width=9.7cm]{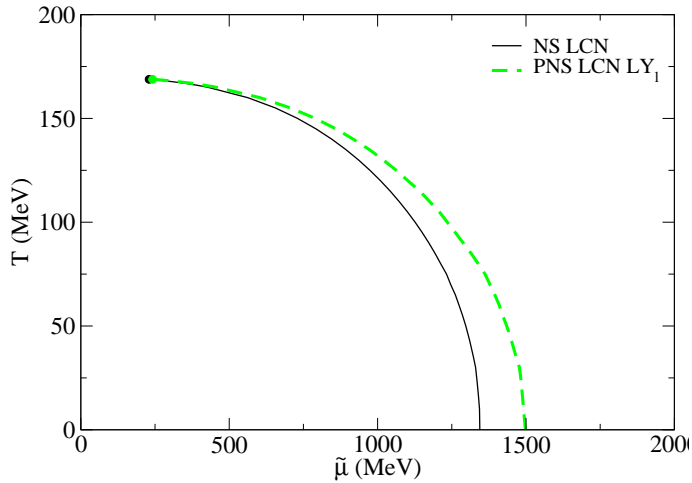}
	\caption{\label{plota}
		The temperature vs$.$ (modified) chemical potential phase diagram for neutron-star matter with locally conserved electric charge and proto-neutron-star matter with locally conserved electric charge and lepton fraction.}
\end{figure}

In the case that electric charge is being conserved globally, we define $\lambda$ as
\begin{eqnarray}
\lambda = \frac{Q_H}{Q_H - Q_Q},
\end{eqnarray}
where $Q_H$ and $Q_Q$ are the electric charge of the hadronic phase and the quark phase, respectively. This equation comes from the constraint of electric charge neutrality of the mixture, namely
\begin{eqnarray}
Q_{mix} = \lambda Q_Q + (1 - \lambda) Q_H = 0.
\end{eqnarray}
In the case that lepton fraction is being conserved globally, we define $\lambda$ as
\begin{eqnarray}
\lambda = \frac{Y_{l,mix} \ n_{B,H}^o - n_{l,H}}{n_{l,Q} - n_{l,H} - Y_{l,mix} \ (n_{B,Q}^o - n_{B,H}^o)}, \nonumber\\
\end{eqnarray}
where the subscripts $H$ and $Q$, again, denote whether the value corresponds to the hadronic phase or the quark phase. This equation comes from the definition of $Y_l$ amongst the phases, namely
\begin{eqnarray}
Y_{l,mix} = \frac{n_{l,mix}}{n_{B,mix}^o},
\end{eqnarray}
where $n_{B,mix}^o = \lambda n_{B,Q}^o + (1 - \lambda) n_{B,H}^o$ and $n_{l,mix} = (n_e + n_\nu)_{mix} = \lambda (n_e + n_\nu)_Q + (1 - \lambda) (n_e + n_\nu)_H$. In the case that electric charge and lepton fraction are \textit{both} conserved globally in proto-neutron-star matter, both values of $\lambda$ are relevant and must be equal for consistency.

In this work, we describe only astrophysical matter (present in different stages of the evolution of stars), in which case net strangeness does not need to be constrained. For this reason, the strange chemical potential is zero and there is no term dependent on strangeness or strange chemical potential in equations (9), (10) and (14). This is not the case for the zero net strangeness matter generated in heavy ion collisions, which is described in detail in Refs.~\cite{Hempel:2009vp,Hempel:2013tfa}.

\section{Results}

\begin{figure}[t!]
	\vspace{3mm}
	\includegraphics[trim={1.4cm 0 0 2.cm},width=9.7cm]{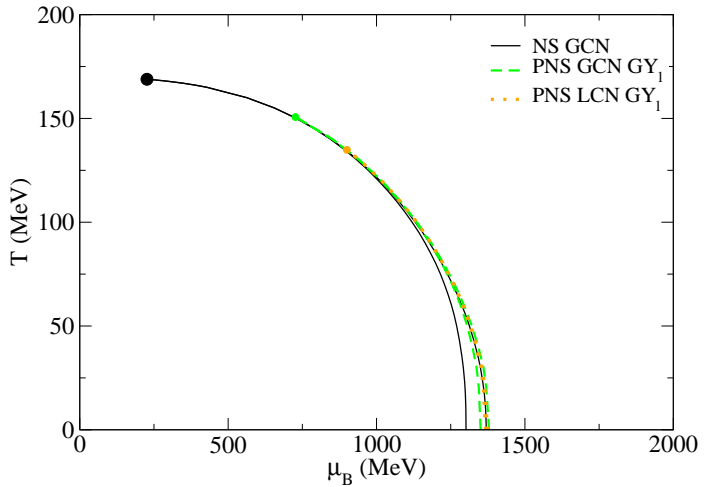}
	\caption{\label{plotb}
		The temperature vs$.$ baryon chemical potential phase diagram for neutron-star matter with globally conserved electric charge, proto-neutron-star matter with locally conserved electric charge and globally conserved lepton fraction, and proto-neutron-star matter with globally conserved electric charge and lepton fraction.}
\end{figure}

The simplest phase diagram we can produce is shown in Fig.~\ref{plota}, for the NS LCN (Id in Ref.~\cite{Hempel:2009vp}) and PNS LCN L$Y_l$ (Ib in Ref.~\cite{Hempel:2009vp}) cases. It shows the deconfinement coexistence lines for NS and PNS matter with all quantities locally conserved. To the left of each coexistence line, we have the hadronic phase, and to the right, the quark phase. The congruent phase transition between these two phases (usually refereed to as Maxwell's construction in astrophysics) is abrupt, as no mixture of phases is possible when all quantities are conserved locally. Physically, local electric charge neutrality is enforced by a possible large surface tension between the phases \cite{Garcia:2013eaa, Lugones:2013ema}. Locally fixed lepton fraction only serves as an academic exercise, as there is no long range force associated with this quantity and, thus, there is no physical reason to expect the lepton fraction to be conserved in a strict, local sense amongst a mixture of phases \cite{Hempel:2009vp}. This scenario is know as a "forced-congruent" case, i.e., lepton fraction is \textit{forced} to be fixed locally as opposed to the more physical case of global conservation. Note, this is the primary motivation for not including a PNS GCN L$Y_l$ case in this paper. The same argument also applies to baryon number density, which is not mentioned as it is always globally conserved. For NS matter, the modified chemical potential is simply equal to the baryon chemical potential in each phase ($\tilde{\mu} = \mu_{B,H} = \mu_{B,Q}$) but, for PNS matter, this is not the case and $\tilde{\mu} = \mu_{B,H} + 0.4 \ \mu_{l,H} = \mu_{B,Q} + 0.4 \ \mu_{l,Q}$.

Still in Fig.~\ref{plota}, note that the coexistence line for PNS matter lies to the right (at a larger modified chemical potential) than the one of NS matter. This difference comes from the way lepton fraction affects each phase differently. In both phases, a larger lepton fraction implies more positive hadrons/quarks (for electric charge neutrality) and, consequently, more isospin-symmetric matter and a softer equation of state. The difference is that the effect is more pronounced in the quark phase, as there are mainly no leptons in this phase in the case of NS matter. This is in agreement, for example, with the conclusions of Ref$.$ \cite{Mariani:2016pcx} performed at fixed entropy.

\begin{figure}[t!]
	\vspace{3mm}
	\includegraphics[trim={1.4cm 0 0 2.cm},width=9.7cm]{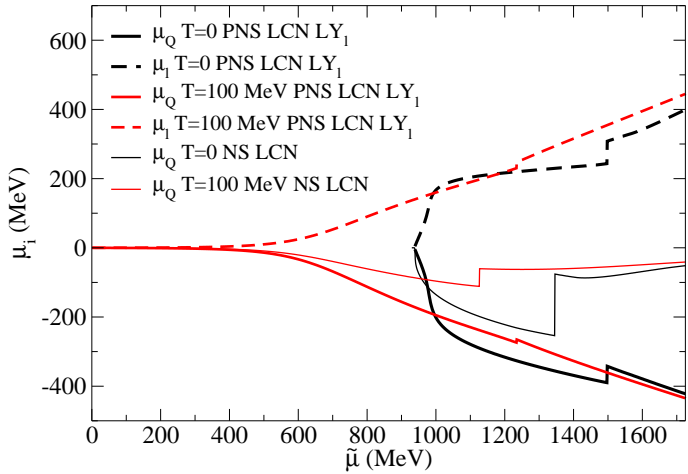}
	\caption{\label{plotg}
		Charged and lepton chemical potentials for neutron-star and proto-neutron-star matter (all charges conserved locally), shown for two different temperatures.}
\end{figure}

Now, Fig.~\ref{plotb} shows the same phase diagram but now allowing quantities to be conserved globally in each case NS GCN, PNS LCN G$Y_l$ (IIb in Ref.~\cite{Hempel:2009vp}), and PNS GCN G$Y_l$ (V in Ref.~\cite{Hempel:2009vp}). When more than one quantity (in addition to baryon chemical potential) is conserved globally, the phase transition becomes non-congruent and a mixture of phases appears (this is usually refereed to as Gibbs' construction in astrophysics), which occupies a region in the phase diagram in Fig.~\ref{plotb}. Note that in all cases the deconfinement and confinement curves are distinct and delimit a region, inside of which the mixture of phases exists. In each case, the left-most curve represents the deconfinement curve where $\lambda=0$ while the right-most curve represents the confinement curve where $\lambda=1$. (Notice that the range in $\mu_B$ of the phase coexistence region in both PNS cases is \textit{extremely} small.) In accordance with the general rules for non-congruent phase transitions, the congruent phase transition coexistence lines from Fig.~\ref{plota} lie between the corresponding deconfinement and confinement lines in all cases when plotted as a function of $\tilde{\mu}$.

In the case of NS matter, the mixture region is large. This fact is related to the large difference in charged chemical potential between the phases (when local charge neutrality is enforced) as shown in Fig.~\ref{plotg} (the full thin black line for $T=0$). In the case of PNS matter, the mixture regions in Fig.~\ref{plotb} are much smaller, as the hadronic and quark phases become more similar within the mixture of phases. This can be seen once more in Fig.~\ref{plotg}, looking at the thick black lines (both full and dashed) for the chemical potentials $\mu_Q$ and $\mu_l$, corresponding to the conserved electric charge and lepton fraction, respectively, for PNS matter.

Still in Fig.~\ref{plotb}, note that the case with \textit{two} globally conserved quantities for PNS's generates a larger mixture region than the case with only one globally conserved quantity. This is quite natural, as the size of the mixture region is related to the number of globally conserved quantities \cite{Hempel:2009vp}.

It is also worth noting the critical point for each case, defined as the point at which first-order phase transitions no longer occur and a smooth crossover appears. Those values are listed in Table~\ref{cps}. The first-order phase transitions are found when, for each baryon chemical potential and temperature, there are multiple metastable solutions in order-parameter space, although only one is truly stable. Notice that the critical points for both coexistence lines in Fig.~\ref{plota} are very close to each other. It is important to note that we believe the accuracy of our numerical calculations is preventing us from going beyond the critical points found in the case of mixtures of phases (Fig.~\ref{plotb}). This happens because the regions containing a mixture of phases (for a fixed temperature) become infinitesimally small in the $T$-$\mu$ plane for large temperatures.

\begin{figure}[t!]
	\vspace{3mm}
	\includegraphics[trim={1.4cm 0 0 2.cm},width=9.7cm]{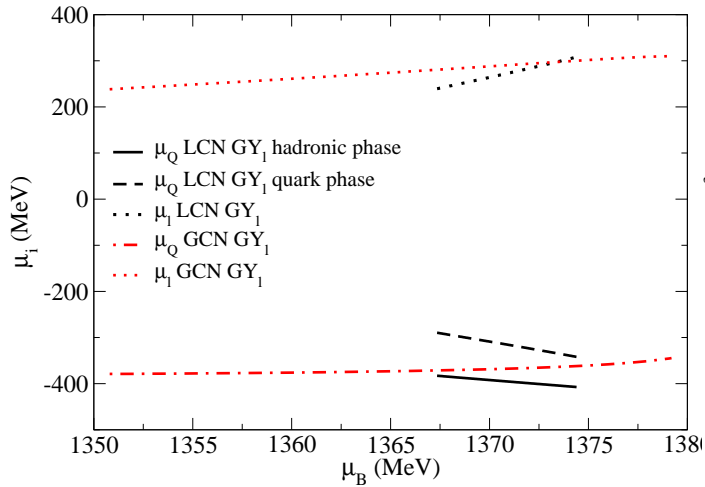}
	\caption{\label{ploth}
		Charged and lepton chemical potentials inside the mixtures of phases in proto-neutron-star matter at zero temperature. Black curves show results for the scenario of locally conserved electric charge and globally conserved lepton fraction and red curves show results for the scenario of globally conserved electric charge and lepton fraction, at $T=0$.}
\end{figure}

\begin{table}
	\caption{\label{cps}
		The critical points, characterized by temperature and modified chemical potential, for all scenarios considered.}
	\begin{ruledtabular}
		\begin{tabular}{ccc}
			NS LCN: & $T_c=168.82$ MeV, & $\tilde{\mu}_c=230.05$ MeV \\
			NS GCN: & $T_c=168.86$ MeV, & $\mu_{B,c}=226.50$ MeV \\
			PNS LCN L$Y_l$: & $T_c=168.84$ MeV, & $\tilde{\mu}_c=241.55$ MeV \\
			PNS LCN G$Y_l$: & $T_c=134.86$ MeV, & $\mu_{B,c}=900.30$ MeV \\
			PNS GCN G$Y_l$: & $T_c=150.65$ MeV, & $\mu_{B,c}=726.65$ MeV \\
		\end{tabular}
	\end{ruledtabular}
\end{table}

Going back to Fig.~\ref{plotg} for a more detailed analysis, we see that in the case that all quantities are conserved locally for NS's (congruent) and PNS's (forced-congruent), there are discontinuities in $\mu_Q$ and $\mu_\nu$ at the phase transitions. This occurs because in congruent cases we do not require, as an equilibrium condition, $\mu_Q$ and $\mu_\nu$ to be equal in both phases. Those "jumps" are smoothed out for larger temperatures (red curves) as the first order phases transition becomes weaker, a physical feature necessary in order to obtain a critical point.

\begin{figure}[t!]
	\vspace{3mm}
	\includegraphics[trim={1.4cm 0 0 2.cm},width=9.7cm]{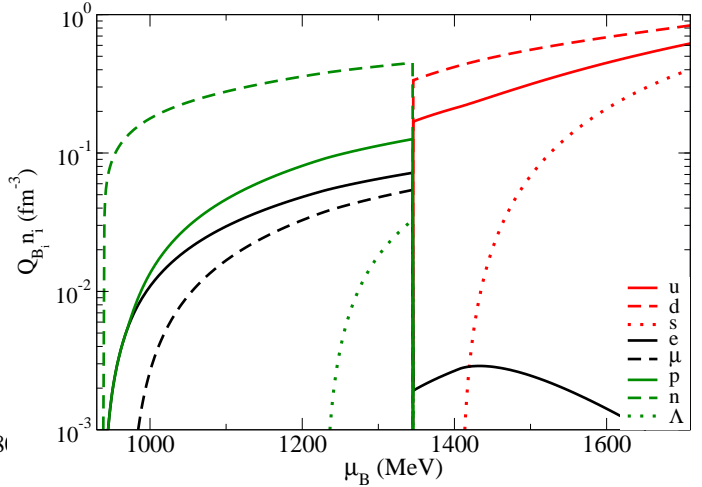}
	\caption{\label{plotc}
		Particle populations for neutron-star matter with locally conserved electric charge, at $T=0$.}
\end{figure}

\begin{figure}[t!]
	\vspace{3mm}
	\includegraphics[trim={1.4cm 0 0 2.cm},width=9.7cm]{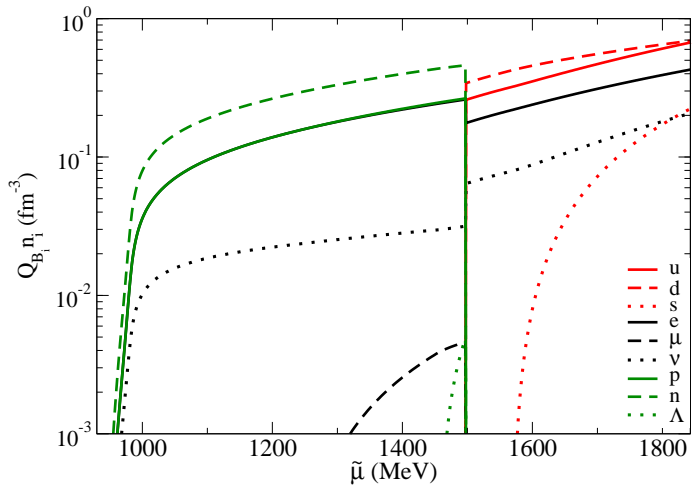}
	\caption{\label{plotc_a}
		Particle populations for proto-neutron-star matter with locally conserved electric charge and lepton fraction, at $T=0$. The curve for electrons mainly overlaps with the curve of protons.}
\end{figure}

Some additional general features of Fig.~\ref{plotg} include the tendency of the lepton chemical potential (dashed lines) to increase as a function of $\tilde{\mu}$. This is quite natural as, when lepton fraction is fixed, the denominator in Eq.~(11) increases with $\tilde{\mu}$, thus forcing the numerator to increase as well. At the phase transition, $\mu_l$ increases as the amount of neutrinos increases and $\mu_\nu = \mu_l$ (particle population plots will be discussed later in Fig.~\ref{plotc} \&~\ref{plotc_a} for NS's and PNS's). As for the charged chemical potential $\mu_Q$ (full lines), it increases in absolute value for NS matter in the hadronic phase as electrons and muons need to balance the increasing amount of positive protons ($\mu_e = \mu_\mu = -\mu_Q$). In the quark phase, $\mu_e$ is lower in absolute value as the down quarks take care of most of the negative contribution to electric charge neutrality. In PNS matter, there are more electrons in the quark phase and the decrease in absolute value for $\mu_Q$ is smaller across the phase transition. Finally, note in Fig.~\ref{plotg} that bulk hadronic matter exists even for small chemical potentials for $T=100$ MeV, while it only starts at the liquid-gas phase transition for $\tilde{\mu}=\mu_B=938$ MeV at zero temperature (before this point, only nucleated matter exists, as opposed to free nucleons).

Similar to Fig.~\ref{plotg}, Fig.~\ref{ploth} features results for PNS matter, allowing for globally conserved quantities (PNS LCN G$Y_l$ and PNS GCN G$Y_l$). As one can see, the condition of local electric charge neutrality radically shrinks the range of $\mu_B$ in the mixture of phases, as already discussed. In the case of local electric charge neutrality, the value of $\mu_Q$ is distinct in each phase, while in the case of globally conserved electric charge, $\mu_Q = \mu_{Q,H} = \mu_{Q,Q}$. Notice how the $\mu_Q$ curve for PNS GCN G$Y_l$ lies between the two $\mu_Q$ curves for PNS LCN G$Y_l$, as expected.

Now we explain in detail the particle populations for each of the cases studied in this work, which should be particularly helpful in distinguishing the differences between the equations of state for NS and PNS matter. In Figs.~\ref{plotc} \&~\ref{plotc_a}, all quantities are locally conserved and, thus, the particle population values change abruptly at the phase transition. In Fig.~\ref{plotc} for NS matter, first there are only neutrons, then protons and electrons, appearing at the same rate (for electric charge neutrality). After that, the muons appear and then the $\Lambda$ hyperons. All other hyperons are suppressed by the phase transition to quark matter at $T=0$. For an example of particle populations for NS's within the CMF model without quarks, see Ref$.$ \cite{Dexheimer:2008ax}. Back to Fig.~\ref{plotc}, in the quark phase, there are down and up quarks with a small amount of electrons, followed by the more massive strange quarks. The y-axis of the figure is in terms of baryon number density, so quark number densities are divided by 3.

In Fig.~\ref{plotc_a}, the population curves for PNS matter are quite different. Besides the obvious change of moving the phase transition to a larger $\tilde{\mu}$ value and the appearance of neutrinos, the ratio of protons to neutrons becomes closer to unity, via a large increase in the number of protons. And in the quark phase, the ratio of up quarks to down quarks also becomes closer to unity, via a large increase in the number of up quarks. As previously mentioned, this occurs as a consequence of fixing both electrons and electron neutrinos in PNS matter. The resulting increase in the number of electrons causes the number of positive hadrons/quarks to increase, as dictated by the condition of electric charge neutrality. This shift to more isospin-symmetric matter contributes to the softening of the equation of state of proto-neutron-star matter, in comparison to neutron-star matter. Notice that the increase in the number of electrons in the hadronic phase leads to a decrease in the number of muons (a result of electric charge neutrality) while the increase in the number of protons in the hadronic phase leads to a decrease in the number of $\Lambda$ hyperons (a result of fixing $Y_l$ and baryon number conservation). This shifting of the appearance of hyperons to larger baryon density values via neutrino trapping, on the other hand, contributes to the stiffening of the equation of state of PNS matter \cite{Nicotra:2005fj, Burgio:2006ed}. Another consequence of PNS matter having more electrons in the quark phase is the suppression of the negative strange quarks. This together with the suppression of hyperons in the hadronic phase (due to the deconfinement phase transition) causes the amount of strangeness to dramatically decrease in both phases of PNS matter.

\begin{figure}[t!]
	\vspace{3mm}
	\includegraphics[trim={1.4cm 0 0 2.cm0},width=9.7cm]{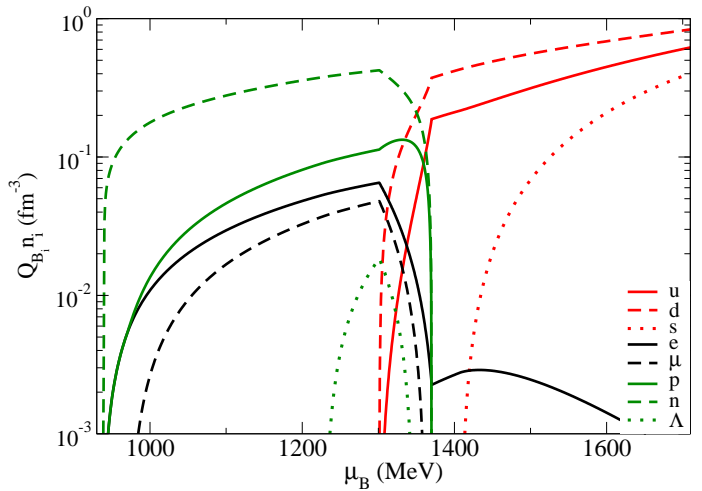}
	\caption{\label{plotd}
		Particle populations for neutron-star matter with globally conserved electric charge, at $T=0$.}
\end{figure}

The large number of down quarks after the deconfinement phase transition means fewer electrons are necessary to ensure electric charge neutrality, thus leading to a drop in the amount of electrons across this phase transition in Figs.~\ref{plotc} \&~\ref{plotc_a}. In PNS matter, an increase in the population of neutrinos in the quark phase accompanies this phenomenon, as the lower number of electrons requires a higher number of electron neutrinos, all in order to maintain $Y_l$ at a value of 0.4.

\begin{figure}[t!]
	\vspace{3mm}
	\includegraphics[trim={1.4cm 0 0 2.cm},width=9.7cm]{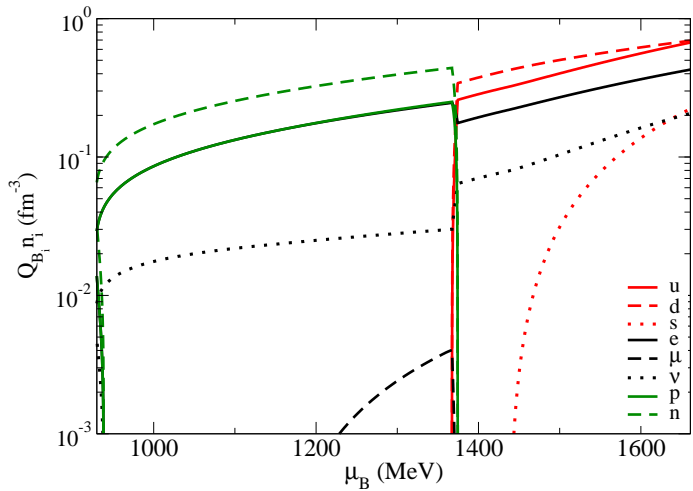}
	\caption{\label{plote}
		Particle populations for proto-neutron-star matter with locally conserved electric charge and globally conserved lepton fraction at, $T=0$. The curve for electrons mainly overlaps with the curve of protons in the hadronic phase.}
\end{figure}

\begin{figure}[t!]
	\vspace{3mm}
	\includegraphics[trim={1.4cm 0 0 2.cm},width=9.7cm]{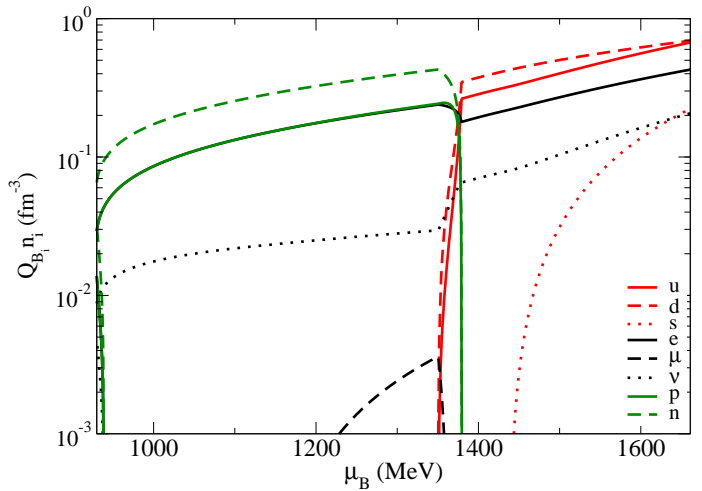}
	\caption{\label{plotf}
		Particle population curves for proto-neutron-star matter with globally conserved electric charge and lepton fraction at, $T=0$. The curve for electrons mainly overlaps with the curve of protons.}
\end{figure}

In contrast, Figs.~\ref{plotd}, ~\ref{plote}, \&~\ref{plotf} feature at least one quantity that is conserved globally and thus, the particle population values change more gradually in non-congruent phase transitions (i.e., this change is made more slowly via the presence of mixtures of phases). More specifically, for NS matter, Fig.~\ref{plotd} is the corresponding version of Fig.~\ref{plotc}, when a mixture of phases with global electric charge neutrality is present. In this case, the quarks appear slowly while the baryons disappear slowly over a range of baryon chemical potential values. The electrons decrease in amount over this range and then, in the quark phase, increase, only to decrease again when the strange quarks appear. In the beginning of the mixture of phases, the quantity of protons increases slightly to balance the negative quarks.

For PNS matter, Figs.~\ref{plote} \&~\ref{plotf} are the corresponding versions of Fig.~\ref{plotc_a} where mixtures of phases with globally fixed lepton fraction are present. In this case, the quarks appear and the baryons disappear over a small range of baryon chemical potential values. The electrons decrease in quantity a bit, the electron neutrinos increase a bit, and the muons disappear over this range. In the case that electric charge neutrality is also conserved globally, the baryon chemical potential range for the mixture of phases is larger (a range of around 30 MeV in Fig.~\ref{plotf} for $T=0$ in comparison to a range of around 8 MeV in Fig.~\ref{plote}).

Due to the difficulty in constraining the equation of state of matter via nuclear physics experiments as we get farther from saturation density, another approach must be taken. Although low energy heavy-ion collision experiments are slowly approaching larger densities (like in FAIR, NICA, and the beam energy scan at RHIC), we are still not in a position to use that data to constrain the equation of state of matter at large densities. For these reasons, we use perturbative QCD (PQCD) to study the behavior of our equation of state at large densities/chemical potentials. In particular, we are going to use results from Ref$.$ \cite{Kurkela:2016was}, where a state-of-the-art three-loop result was derived for the pressure of deconfined quark matter, valid for all values of temperature and density in the asymptotically high energy regime. This calculation employed resummations provided by dimensional reduction \cite{Braaten:1995cm} and Hard Thermal Loop \cite{Braaten:1991gm} effective theories to account for the contributions of both static and non-static long-distance gluon fields, and added to this a contribution from the perturbative hard field modes, obtained from Ref$.$ \cite{Vuorinen:2003fs}. The uncertainty of the result can be roughly estimated from its dependence on the renormalization scale, which is conventionally varied by a factor of two around a central value to obtain a band of viable equations of state, although other factors such as higher-order terms in the perturbative expansion, truly non-perturbative contributions, etc.\ are not accounted for.

While a very versatile result that is immediately available both in and out of beta equilibrium, the perturbative pressure can only be trusted at relatively high energy densities. This stems not only from the running of the gauge coupling, but also from the fact that quark masses have been set to zero in the perturbative computation of Ref$.$ \cite{Kurkela:2016was}, implying that the strange quark mass has to be negligible in comparison with either the temperature or the baryon chemical potential. According to the criterion used e.g.~in Refs$.$ \cite{Kurkela:2014vha,Annala:2017llu}, at $T=0$ the perturbative results can be considered to be reasonably trustworthy from ca. $\mu_B=2.4$ GeV onwards, where their relative uncertainty is ca.~$\pm 23\%$. Recently, Ref$.$ \cite{Jimenez:2017fax} presented PQCD results with charge neutrality and fixed lepton fraction but only for zero temperature. Here, we present for the first time PQCD results suited for for proto-neutron-star conditions at relevant finite temperatures.

\begin{figure}[t!]
	\vspace{3mm}
	\includegraphics[trim={1.4cm 0 0 2.cm},width=9.7cm]{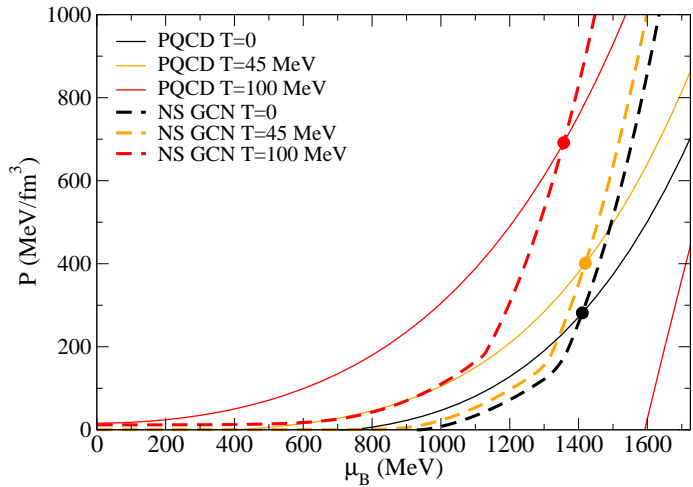}
	\caption{\label{plotj}
		Comparison of our results for neutron-star matter (with globally conserved electric charge) with results provided by PQCD, shown for several temperatures (Ref$.$ \cite{Kurkela:2016was}). Except for the largest temperature, the lower edge of the PQCD regions lie to the right of the figure bounds.}
\end{figure}

Fig.~\ref{plotj} shows our equation of state for NS matter with global charge neutrality (dashed lines) for several temperatures relevant for cold neutron stars, proto-neutron stars, and neutron-star mergers. In the same figure, we plot the PQCD results (full lines) for the same conditions (chemical equilibrium and electric charge neutrality). For the largest temperature ($T=100$ MeV), the lower PQCD bound appears in the figure, but for the two lower temperatures, their lower bounds lie to the right of the figure. The colored points indicate the baryon chemical potential/pressure beyond which our equation of state starts to disagree (by becoming stiffer) with the PQCD results. The corresponding baryon chemical potentials and pressures that characterize these points are shown in Table~\ref{limits}. Note that PQCD calculations cannot be applied below baryon chemical potentials at which the hadrons are expected to be present, as they only contain quark degrees of freedom. Remember that in our formalism, as temperature goes up, hadrons start to appear in the so-called quark phase. At $T=0$, our equation of state predicts a neutron star with a central baryon chemical potential of $1319$ MeV, well below the PQCD limiting value. It is worth mentioning once more, that varying the renormalization scale by a larger value than what was used in the PQCD results presented here (or including other effects) would increase the size of their region accordingly.

\begin{figure}[t!]
	\vspace{3mm}
	\includegraphics[trim={1.4cm 0 0 2.cm},width=9.7cm]{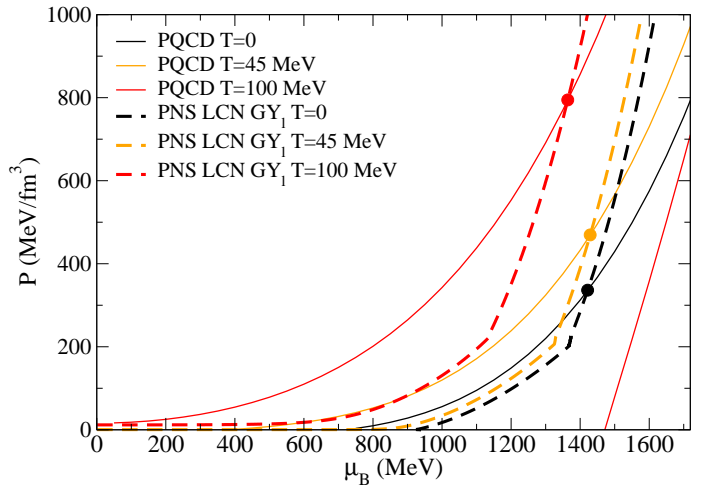}
	\caption{\label{plotk}
		Comparison of our results for proto-neutron-star matter (with locally conserved electric charge and globally conserved lepton fraction) with results provided by PQCD, shown for several temperatures (Ref$.$ \cite{Kurkela:2016was}). Except for the largest temperature, the lower edge of the PQCD regions lie to the right of the figure bounds.}
\end{figure}

Fig.~\ref{plotk} is analogous to Fig.~\ref{plotj}, but for PNS matter. In this case, we agree with PQCD up until relatively larger baryon chemical potentials. This is expected, as both we and PQCD calculations treat the leptons in exactly the same way, as a free gas. Again, the corresponding baryon chemical potentials and pressures that characterize the colored points beyond which we do not agree with PQCD results are shown in Table~\ref{limits}. Finally, note that our curves are shown for the PNS LCN G$Y_l$ case. Otherwise being identical, the PNS GCN G$Y_l$ case would only smooth out the "kink" at the phase transition and would not change the position of the colored points. For PNS structure calculations, the temperature cannot be fixed throughout each star, nevertheless, a rough estimate predicts within the CMF model stars with a central baryon chemical potential around $1330$ MeV, again well below the PQCD limiting value. This will be studied in detail in another publication.

\begin{table}
	\caption{\label{limits}
		The pressure (measured in MeV/fm$^3$), baryon chemical potential (measured in MeV), and corresponding baryon number density values after which our results no longer lie between the limits provided by PQCD results, for neutron-star matter (top) and proto-neutron-star matter (bottom) at several temperatures. Note that we use for saturation density $\rho_0=0.15$ fm$^{-3}$.}
	\begin{ruledtabular}
		\begin{tabular}{cccc}
			$T=0$: & $P=281.75$, & $\mu_B=1411.04$, & $\rho_B/\rho_0=14.93$ \\
			$T=45$ MeV: & $P=401.10$, & $\mu_B=1419.76$, & $\rho_B/\rho_0=17.27$ \\
			$T=100$ MeV: & $P=691.47$, & $\mu_B=1356.87$, & $\rho_B/\rho_0=20.20$ \\
		\end{tabular}
		\begin{tabular}{cccc}
			$T=0$: & $P=336.27$, & $\mu_B=1421.69$, & $\rho_B/\rho_0=15.67$ \\
			$T=45$ MeV: & $P=469.42$, & $\mu_B=1429.09$, & $\rho_B/\rho_0=17.80$ \\
			$T=100$ MeV: & $P=794.52$, & $\mu_B=1364.08$, & $\rho_B/\rho_0=20.60$ \\
		\end{tabular}
	\end{ruledtabular}
\end{table}

\begin{figure}[t!]
	\vspace{3mm}
	\includegraphics[trim={1.4cm 0 0 2.cm},width=9.7cm]{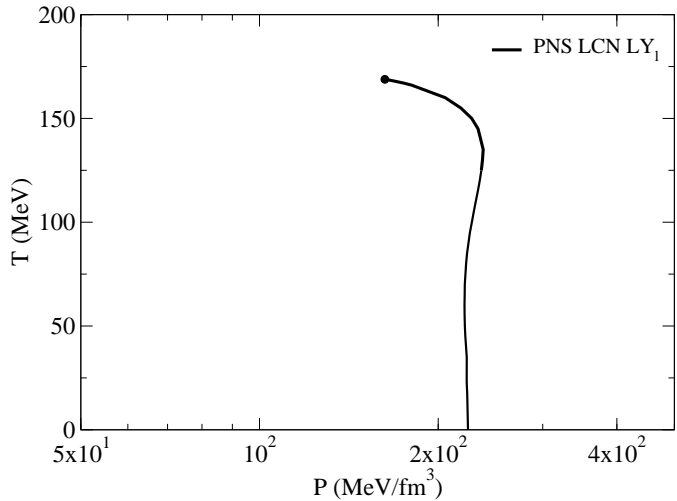}
	\caption{\label{plotn}
		The temperature vs$.$ pressure phase diagram for proto-neutron-star matter with locally conserved electric charge and lepton fraction.}
\end{figure}

In preparation for analyzing Fig.~\ref{plotn} (a phase diagram in the $T$-$P$ plane), let us discuss the Clausius-Clapeyron relation, which relates $dP/dT$ to the difference in the entropy per baryon (or entropy density per baryon density) $S_B$ between two coexisting phases and the difference in the baryon number density between these phases:
\begin{eqnarray}
\frac{dP}{dT} = \frac{S_{B,I} - S_{B,II}}{1/n_{B,I} - 1/n_{B,II}}.
\end{eqnarray}
Now, in the case of the liquid-gas phase transition (general or nuclear), $dP/dT$ is positive and, thus, correctly implies that $S_{B,G} > S_{B,L}$. In the case of the deconfinement phase transition in heavy-ion matter \cite{Hempel:2013tfa, Hempel:2017hsx} and neutron star matter \cite{Bombaci:2009jt}, $dP/dT$ was shown to be instead \textit{negative}. This then implies that $S_{B,Q} > S_{B,H}$, which is quite natural as the quarks contain extra color degrees of freedom. Such a case can lead to unexpected thermodynamic properties \cite{Yudin:2015cva, Hempel:2015vlg}. But, as we can see for the PNS LCN L$Y_l$ case in Fig.~\ref{plotn}, in the direction of increasing temperature, $dP/dT$ starts off as being positive at intermediate temperatures and then becomes negative.

This result is not surprising, as requiring electric charge neutrality and fixing $Y_l$ significantly modifies the kinds of degrees of freedom in the quark phase by increasing the amount of leptons, which do not contain color degrees of freedom. Numerically, the total entropy density over baryon density flips sign across the coexistence line around $T=135$ MeV when the increase in the $\Phi$ contirbution $s_\Phi/n_B$ becomes larger than the decrease in the fermionic contribution $(s_B + s_l)/n_B$.

\begin{figure}[t!]
	\vspace{3mm}
	\includegraphics[trim={1.4cm 0 0 2.cm},width=9.7cm]{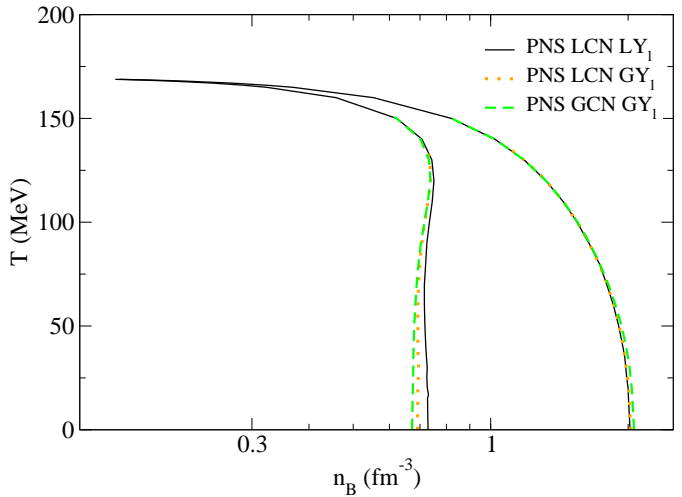}
	\caption{\label{ploto}
		The temperature vs$.$ number density phase diagram for proto-neutron-star matter with locally conserved electric charge and lepton fraction, locally conserved electric charge and globally conserved lepton fraction, and globally conserved electric charge and lepton fraction.}
\end{figure}

In Fig.~\ref{ploto}, we show the phase diagram in the $T$-$n_B$ plane for all three PNS cases. In the case of locally fixed $Y_l$, there is a jump for each temperature from the hadronic phase value $n_{B,H}$ to the quark phase value $n_{B,Q}$, as this is a congruent phase transition. In the PNS LCN G$Y_l$ case there is a mixture of phases, each phase having a different $\mu_Q$ value (locally charge neutral) but the same $\mu_l$ value (black lines in Fig.~\ref{ploth}) in a way that $n_B$ increases continuously for a given temperature. In the PNS GCN G$Y_l$ case, $\mu_Q$ and $\mu_l$ are both the same in each phase. At this point it is important to note again that we believe the accuracy of our numerical calculations is preventing us from going beyond the critical points found for the mixture of phases for both of the PNS cases with at least one globally conserved quantity. In this case, the PNS LCN G$Y_l$ and PNS GCN G$Y_l$ lines would continue to go up and would eventually merge in Fig.~\ref{ploto}. Nevertheless, by comparing Fig.~\ref{ploto} with Fig$.$ 2 from Ref$.$ \cite{Dexheimer:2012qk} and Ref$.$ \cite{Fischer:2011zj}, it becomes clear that a fixed, large $Y_l$ pushes the deconfinement phase transtion to larger baryon number densities.

\begin{figure}[t!]
	\vspace{3mm}
	\includegraphics[trim={1.4cm 0 0 2.cm},width=9.7cm]{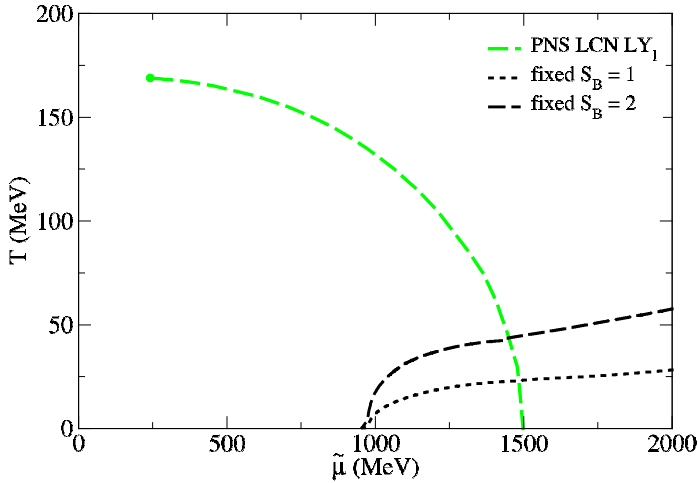}
	\caption{\label{plotp}
		The temperature vs$.$ modified chemical potential phase diagram for proto-neutron-star matter with locally conserved electric charge and lepton fraction. Two example trajectories show the temperature corresponding to a fixed value of entropy density per baryon density ($S_B=s/n_B$) for each modified chemical potential.}
\end{figure}

Finally, Fig.~\ref{plotp} shows again the PNS matter curve from Fig.~\ref{plota} together with two example trajectories of the temperature inside a PNS in which entropy density per baryon density is fixed, in the simple case where all quantities are conserved locally. Such treatment results in a small jump in temperature across the phase transition, not following the condition of thermal equilibrium. As explained in section 3D of Ref$.$ \cite{Hempel:2009vp}, this is not the correct treatment and a mixture of phases must be accounted for. We are currently working on an extended treatment for the case in which the entropy is fixed across the phase transition, instead of temperature. In any case, these jumps are of about 1 MeV in the cases shown in Fig.~\ref{plotp}.

\section{Conclusions and Outlook}

The relativistic, mean-field formalism based on the CMF model described in this work is ideal for describing the features of the phase diagram, as it effectively describes both hadronic and quark matter. Note that this is the usual procedure when studying the nuclear liquid-gas phase transition \cite{Landau:1969}. This allows us to study congruent phase transitions (where there is no phase coexistence and the phases are distinctly separated) and non-congruent phase transitions (where there is a phase coexistence of two or more macroscopic phases with different chemical compositions), the latter featuring a mixture of phases. Note that non-congruent deconfinement phase transitions have somehow been studied in the past, even with fixed lepton fraction but using the bag model \cite{Yasutake:2012dw}, and without the introduction of the modified chemical potential $\tilde{\mu}$.

In this work, we built on our previous work by investigating the effects of neutrino trapping and the consequent consideration of fixing lepton fraction $Y_l$ in the phase transition associated with quark deconfinement. This provided us with a new conserved quantity (in addition to charge neutrality and baryon number conservation). This in turn suppressed the hyperons and pushed the deconfinement phase transition to higher chemical potentials. A new modified chemical potential was introduced and simple coexistence curves were shifted to larger chemical potentials due to the fixing of $Y_l$, which made the quark matter equation of state softer than the hadronic one. Different phase diagrams and particle population figures were shown and discussed and comparisons with our previous works were drawn. Mixtures of phases with different globally conserved quantities were also obtained. As a result, we found that proto-neutron-star matter possess much smaller mixtures of phases than those of neutron-star matter (i.e., they extend through much smaller ranges of chemical potentials and smaller ranges of densities). This is a very optimistic result as these mixtures of phases, which could possibly disguise a signal for deconfinement in proto-neutron stars, are reduced. This information will be particularly helpful when it comes to interpreting signals from supernova events observed in the future.

Additionally, for the first time a thorough analysis of the behavior of the CMF model in comparison with PQCD was performed at large densities and temperatures. It was shown that the model is in agreement with PQCD results calculated for neutron and proto-neutron star conditions for all ranges that can exist inside compact stars. The results from PQCD presented here are in the form of a band, which includes some but not all of the uncertainties in their calculations.

In the future, we will be focusing on the consequences of mixtures of phases for the macroscopic structure of proto-neutron stars with fixed entropy, together with rotation and magnetic field effects. This has been previously shown in Ref$.$ \cite{Franzon:2016iai} where, for proto-neutron stars containing only hadrons, the space anisotropy created by the magnetic field also creates an anisotropy in the amount of neutrinos in the star. We would like to investigate this again in the presence of phase transitions. In addition, we intend to study phase transitions and mixtures of phases in the context of neutron stars mergers. More specifically, we want for example to check if phase transitions can change relations between neutron star radii and tidal deformabilities, such as the one found in Ref$.$ \cite{Most:2018hfd}.

The authors would like to thank Matthias Hempel and Stefan Schramm for helpful discussions. We would also like to thank Aleksi Vuorinen for providing the PQCD results employed in this paper.

\bibliographystyle{apsrev4-1}
\bibliography{paper}

\begin{thebibliography}{66}%
\makeatletter
\providecommand \@ifxundefined [1]{%
 \@ifx{#1\undefined}
}%
\providecommand \@ifnum [1]{%
 \ifnum #1\expandafter \@firstoftwo
 \else \expandafter \@secondoftwo
 \fi
}%
\providecommand \@ifx [1]{%
 \ifx #1\expandafter \@firstoftwo
 \else \expandafter \@secondoftwo
 \fi
}%
\providecommand \natexlab [1]{#1}%
\providecommand \enquote  [1]{``#1''}%
\providecommand \bibnamefont  [1]{#1}%
\providecommand \bibfnamefont [1]{#1}%
\providecommand \citenamefont [1]{#1}%
\providecommand \href@noop [0]{\@secondoftwo}%
\providecommand \href [0]{\begingroup \@sanitize@url \@href}%
\providecommand \@href[1]{\@@startlink{#1}\@@href}%
\providecommand \@@href[1]{\endgroup#1\@@endlink}%
\providecommand \@sanitize@url [0]{\catcode `\\12\catcode `\$12\catcode
  `\&12\catcode `\#12\catcode `\^12\catcode `\_12\catcode `\%12\relax}%
\providecommand \@@startlink[1]{}%
\providecommand \@@endlink[0]{}%
\providecommand \url  [0]{\begingroup\@sanitize@url \@url }%
\providecommand \@url [1]{\endgroup\@href {#1}{\urlprefix }}%
\providecommand \urlprefix  [0]{URL }%
\providecommand \Eprint [0]{\href }%
\providecommand \doibase [0]{http://dx.doi.org/}%
\providecommand \selectlanguage [0]{\@gobble}%
\providecommand \bibinfo  [0]{\@secondoftwo}%
\providecommand \bibfield  [0]{\@secondoftwo}%
\providecommand \translation [1]{[#1]}%
\providecommand \BibitemOpen [0]{}%
\providecommand \bibitemStop [0]{}%
\providecommand \bibitemNoStop [0]{.\EOS\space}%
\providecommand \EOS [0]{\spacefactor3000\relax}%
\providecommand \BibitemShut  [1]{\csname bibitem#1\endcsname}%
\let\auto@bib@innerbib\@empty
\bibitem [{\citenamefont {Lukacs}\ \emph {et~al.}(1987)\citenamefont {Lukacs},
  \citenamefont {Zimanyi},\ and\ \citenamefont {Balazs}}]{Lukacs:1986hu}%
  \BibitemOpen
  \bibfield  {author} {\bibinfo {author} {\bibfnamefont {B.}~\bibnamefont
  {Lukacs}}, \bibinfo {author} {\bibfnamefont {J.}~\bibnamefont {Zimanyi}}, \
  and\ \bibinfo {author} {\bibfnamefont {N.~L.}\ \bibnamefont {Balazs}},\
  }\href {\doibase 10.1016/0370-2693(87)91411-0} {\bibfield  {journal}
  {\bibinfo  {journal} {Phys. Lett.}\ }\textbf {\bibinfo {volume} {B183}},\
  \bibinfo {pages} {27} (\bibinfo {year} {1987})}\BibitemShut {NoStop}%
\bibitem [{\citenamefont {Heinz}\ \emph {et~al.}(1987)\citenamefont {Heinz},
  \citenamefont {Lee},\ and\ \citenamefont {Rhoades-Brown}}]{Heinz:1987sj}%
  \BibitemOpen
  \bibfield  {author} {\bibinfo {author} {\bibfnamefont {U.~W.}\ \bibnamefont
  {Heinz}}, \bibinfo {author} {\bibfnamefont {K.~S.}\ \bibnamefont {Lee}}, \
  and\ \bibinfo {author} {\bibfnamefont {M.~J.}\ \bibnamefont
  {Rhoades-Brown}},\ }\href {\doibase 10.1142/S0217732387000197} {\bibfield
  {journal} {\bibinfo  {journal} {Mod. Phys. Lett.}\ }\textbf {\bibinfo
  {volume} {A2}},\ \bibinfo {pages} {153} (\bibinfo {year} {1987})}\BibitemShut
  {NoStop}%
\bibitem [{\citenamefont {Glendenning}(1992)}]{Glendenning:1992vb}%
  \BibitemOpen
  \bibfield  {author} {\bibinfo {author} {\bibfnamefont {N.~K.}\ \bibnamefont
  {Glendenning}},\ }\href {\doibase 10.1103/PhysRevD.46.1274} {\bibfield
  {journal} {\bibinfo  {journal} {Phys. Rev.}\ }\textbf {\bibinfo {volume}
  {D46}},\ \bibinfo {pages} {1274} (\bibinfo {year} {1992})}\BibitemShut
  {NoStop}%
\bibitem [{\citenamefont {Hempel}\ \emph {et~al.}(2013)\citenamefont {Hempel},
  \citenamefont {Dexheimer}, \citenamefont {Schramm},\ and\ \citenamefont
  {Iosilevskiy}}]{Hempel:2013tfa}%
  \BibitemOpen
  \bibfield  {author} {\bibinfo {author} {\bibfnamefont {M.}~\bibnamefont
  {Hempel}}, \bibinfo {author} {\bibfnamefont {V.}~\bibnamefont {Dexheimer}},
  \bibinfo {author} {\bibfnamefont {S.}~\bibnamefont {Schramm}}, \ and\
  \bibinfo {author} {\bibfnamefont {I.}~\bibnamefont {Iosilevskiy}},\ }\href
  {\doibase 10.1103/PhysRevC.88.014906} {\bibfield  {journal} {\bibinfo
  {journal} {Phys. Rev.}\ }\textbf {\bibinfo {volume} {C88}},\ \bibinfo {pages}
  {014906} (\bibinfo {year} {2013})}\BibitemShut {NoStop}%
\bibitem [{\citenamefont {Prakash}\ \emph {et~al.}(1997)\citenamefont
  {Prakash}, \citenamefont {Bombaci}, \citenamefont {Prakash}, \citenamefont
  {Ellis}, \citenamefont {Lattimer},\ and\ \citenamefont
  {Knorren}}]{Prakash:1996xs}%
  \BibitemOpen
  \bibfield  {author} {\bibinfo {author} {\bibfnamefont {M.}~\bibnamefont
  {Prakash}}, \bibinfo {author} {\bibfnamefont {I.}~\bibnamefont {Bombaci}},
  \bibinfo {author} {\bibfnamefont {M.}~\bibnamefont {Prakash}}, \bibinfo
  {author} {\bibfnamefont {P.~J.}\ \bibnamefont {Ellis}}, \bibinfo {author}
  {\bibfnamefont {J.~M.}\ \bibnamefont {Lattimer}}, \ and\ \bibinfo {author}
  {\bibfnamefont {R.}~\bibnamefont {Knorren}},\ }\href {\doibase
  10.1016/S0370-1573(96)00023-3} {\bibfield  {journal} {\bibinfo  {journal}
  {Phys. Rept.}\ }\textbf {\bibinfo {volume} {280}},\ \bibinfo {pages} {1}
  (\bibinfo {year} {1997})}\BibitemShut {NoStop}%
\bibitem [{\citenamefont {Reddy}\ \emph {et~al.}(1998)\citenamefont {Reddy},
  \citenamefont {Prakash},\ and\ \citenamefont {Lattimer}}]{Reddy:1997yr}%
  \BibitemOpen
  \bibfield  {author} {\bibinfo {author} {\bibfnamefont {S.}~\bibnamefont
  {Reddy}}, \bibinfo {author} {\bibfnamefont {M.}~\bibnamefont {Prakash}}, \
  and\ \bibinfo {author} {\bibfnamefont {J.~M.}\ \bibnamefont {Lattimer}},\
  }\href {\doibase 10.1103/PhysRevD.58.013009} {\bibfield  {journal} {\bibinfo
  {journal} {Phys. Rev.}\ }\textbf {\bibinfo {volume} {D58}},\ \bibinfo {pages}
  {013009} (\bibinfo {year} {1998})}\BibitemShut {NoStop}%
\bibitem [{\citenamefont {Shen}\ \emph {et~al.}(2003)\citenamefont {Shen},
  \citenamefont {Lombardo}, \citenamefont {Van~Giai},\ and\ \citenamefont
  {Zuo}}]{Shen:2003ih}%
  \BibitemOpen
  \bibfield  {author} {\bibinfo {author} {\bibfnamefont {C.}~\bibnamefont
  {Shen}}, \bibinfo {author} {\bibfnamefont {U.}~\bibnamefont {Lombardo}},
  \bibinfo {author} {\bibfnamefont {N.}~\bibnamefont {Van~Giai}}, \ and\
  \bibinfo {author} {\bibfnamefont {W.}~\bibnamefont {Zuo}},\ }\href {\doibase
  10.1103/PhysRevC.68.055802} {\bibfield  {journal} {\bibinfo  {journal} {Phys.
  Rev.}\ }\textbf {\bibinfo {volume} {C68}},\ \bibinfo {pages} {055802}
  (\bibinfo {year} {2003})}\BibitemShut {NoStop}%
\bibitem [{\citenamefont {Pastore}\ \emph {et~al.}(2014)\citenamefont
  {Pastore}, \citenamefont {Martini}, \citenamefont {Davesne}, \citenamefont
  {Navarro}, \citenamefont {Goriely},\ and\ \citenamefont
  {Chamel}}]{Pastore:2014yua}%
  \BibitemOpen
  \bibfield  {author} {\bibinfo {author} {\bibfnamefont {A.}~\bibnamefont
  {Pastore}}, \bibinfo {author} {\bibfnamefont {M.}~\bibnamefont {Martini}},
  \bibinfo {author} {\bibfnamefont {D.}~\bibnamefont {Davesne}}, \bibinfo
  {author} {\bibfnamefont {J.}~\bibnamefont {Navarro}}, \bibinfo {author}
  {\bibfnamefont {S.}~\bibnamefont {Goriely}}, \ and\ \bibinfo {author}
  {\bibfnamefont {N.}~\bibnamefont {Chamel}},\ }\href {\doibase
  10.1103/PhysRevC.90.025804} {\bibfield  {journal} {\bibinfo  {journal} {Phys.
  Rev.}\ }\textbf {\bibinfo {volume} {C90}},\ \bibinfo {pages} {025804}
  (\bibinfo {year} {2014})}\BibitemShut {NoStop}%
\bibitem [{\citenamefont {Carter}(2002)}]{Carter:2001ji}%
  \BibitemOpen
  \bibfield  {author} {\bibinfo {author} {\bibfnamefont {G.~W.}\ \bibnamefont
  {Carter}},\ }\bibfield  {booktitle} {\emph {\bibinfo {booktitle} {{Compact
  stars in the QCD phase diagram. Proceedings, Conference, Copenhagen, Denmark,
  August 15-18, 2001}}},\ }\href@noop {} {\bibfield  {journal} {\bibinfo
  {journal} {eConf}\ }\textbf {\bibinfo {volume} {C010815}},\ \bibinfo {pages}
  {149} (\bibinfo {year} {2002})}\BibitemShut {NoStop}%
\bibitem [{\citenamefont {Constantinou}\ \emph {et~al.}(2014)\citenamefont
  {Constantinou}, \citenamefont {Muccioli}, \citenamefont {Prakash},\ and\
  \citenamefont {Lattimer}}]{Constantinou:2014hha}%
  \BibitemOpen
  \bibfield  {author} {\bibinfo {author} {\bibfnamefont {C.}~\bibnamefont
  {Constantinou}}, \bibinfo {author} {\bibfnamefont {B.}~\bibnamefont
  {Muccioli}}, \bibinfo {author} {\bibfnamefont {M.}~\bibnamefont {Prakash}}, \
  and\ \bibinfo {author} {\bibfnamefont {J.~M.}\ \bibnamefont {Lattimer}},\
  }\href {\doibase 10.1103/PhysRevC.89.065802} {\bibfield  {journal} {\bibinfo
  {journal} {Phys. Rev.}\ }\textbf {\bibinfo {volume} {C89}},\ \bibinfo {pages}
  {065802} (\bibinfo {year} {2014})}\BibitemShut {NoStop}%
\bibitem [{\citenamefont {Du}\ \emph {et~al.}(2018)\citenamefont {Du},
  \citenamefont {Steiner},\ and\ \citenamefont {Holt}}]{Du:2018vyp}%
  \BibitemOpen
  \bibfield  {author} {\bibinfo {author} {\bibfnamefont {X.}~\bibnamefont
  {Du}}, \bibinfo {author} {\bibfnamefont {A.~W.}\ \bibnamefont {Steiner}}, \
  and\ \bibinfo {author} {\bibfnamefont {J.~W.}\ \bibnamefont {Holt}},\
  }\href@noop {} {\  (\bibinfo {year} {2018})},\ \Eprint
  {http://arxiv.org/abs/1802.09710} {arXiv:1802.09710 [nucl-th]} \BibitemShut
  {NoStop}%
\bibitem [{\citenamefont {Nakazato}\ \emph {et~al.}(2008)\citenamefont
  {Nakazato}, \citenamefont {Sumiyoshi},\ and\ \citenamefont
  {Yamada}}]{Nakazato:2008su}%
  \BibitemOpen
  \bibfield  {author} {\bibinfo {author} {\bibfnamefont {K.}~\bibnamefont
  {Nakazato}}, \bibinfo {author} {\bibfnamefont {K.}~\bibnamefont {Sumiyoshi}},
  \ and\ \bibinfo {author} {\bibfnamefont {S.}~\bibnamefont {Yamada}},\ }\href
  {\doibase 10.1103/PhysRevD.77.103006} {\bibfield  {journal} {\bibinfo
  {journal} {Phys. Rev.}\ }\textbf {\bibinfo {volume} {D77}},\ \bibinfo {pages}
  {103006} (\bibinfo {year} {2008})}\BibitemShut {NoStop}%
\bibitem [{\citenamefont {Sagert}\ \emph {et~al.}(2009)\citenamefont {Sagert},
  \citenamefont {Fischer}, \citenamefont {Hempel}, \citenamefont {Pagliara},
  \citenamefont {Schaffner-Bielich}, \citenamefont {Mezzacappa}, \citenamefont
  {Thielemann},\ and\ \citenamefont {Liebendorfer}}]{Sagert:2008ka}%
  \BibitemOpen
  \bibfield  {author} {\bibinfo {author} {\bibfnamefont {I.}~\bibnamefont
  {Sagert}}, \bibinfo {author} {\bibfnamefont {T.}~\bibnamefont {Fischer}},
  \bibinfo {author} {\bibfnamefont {M.}~\bibnamefont {Hempel}}, \bibinfo
  {author} {\bibfnamefont {G.}~\bibnamefont {Pagliara}}, \bibinfo {author}
  {\bibfnamefont {J.}~\bibnamefont {Schaffner-Bielich}}, \bibinfo {author}
  {\bibfnamefont {A.}~\bibnamefont {Mezzacappa}}, \bibinfo {author}
  {\bibfnamefont {F.~K.}\ \bibnamefont {Thielemann}}, \ and\ \bibinfo {author}
  {\bibfnamefont {M.}~\bibnamefont {Liebendorfer}},\ }\href {\doibase
  10.1103/PhysRevLett.102.081101} {\bibfield  {journal} {\bibinfo  {journal}
  {Phys. Rev. Lett.}\ }\textbf {\bibinfo {volume} {102}},\ \bibinfo {pages}
  {081101} (\bibinfo {year} {2009})}\BibitemShut {NoStop}%
\bibitem [{\citenamefont {Nakazato}\ \emph {et~al.}(2010)\citenamefont
  {Nakazato}, \citenamefont {Sumiyoshi},\ and\ \citenamefont
  {Yamada}}]{Nakazato:2010ue}%
  \BibitemOpen
  \bibfield  {author} {\bibinfo {author} {\bibfnamefont {K.}~\bibnamefont
  {Nakazato}}, \bibinfo {author} {\bibfnamefont {K.}~\bibnamefont {Sumiyoshi}},
  \ and\ \bibinfo {author} {\bibfnamefont {S.}~\bibnamefont {Yamada}},\ }\href
  {\doibase 10.1088/0004-637X/721/2/1284} {\bibfield  {journal} {\bibinfo
  {journal} {Astrophys. J.}\ }\textbf {\bibinfo {volume} {721}},\ \bibinfo
  {pages} {1284} (\bibinfo {year} {2010})}\BibitemShut {NoStop}%
\bibitem [{\citenamefont {Fischer}\ \emph {et~al.}(2011)\citenamefont
  {Fischer}, \citenamefont {Sagert}, \citenamefont {Pagliara}, \citenamefont
  {Hempel}, \citenamefont {Schaffner-Bielich}, \citenamefont {Rauscher},
  \citenamefont {Thielemann}, \citenamefont {Kappeli}, \citenamefont
  {Martinez-Pinedo},\ and\ \citenamefont {Liebendorfer}}]{Fischer:2010wp}%
  \BibitemOpen
  \bibfield  {author} {\bibinfo {author} {\bibfnamefont {T.}~\bibnamefont
  {Fischer}}, \bibinfo {author} {\bibfnamefont {I.}~\bibnamefont {Sagert}},
  \bibinfo {author} {\bibfnamefont {G.}~\bibnamefont {Pagliara}}, \bibinfo
  {author} {\bibfnamefont {M.}~\bibnamefont {Hempel}}, \bibinfo {author}
  {\bibfnamefont {J.}~\bibnamefont {Schaffner-Bielich}}, \bibinfo {author}
  {\bibfnamefont {T.}~\bibnamefont {Rauscher}}, \bibinfo {author}
  {\bibfnamefont {F.~K.}\ \bibnamefont {Thielemann}}, \bibinfo {author}
  {\bibfnamefont {R.}~\bibnamefont {Kappeli}}, \bibinfo {author} {\bibfnamefont
  {G.}~\bibnamefont {Martinez-Pinedo}}, \ and\ \bibinfo {author} {\bibfnamefont
  {M.}~\bibnamefont {Liebendorfer}},\ }\href {\doibase
  10.1088/0067-0049/194/2/39} {\bibfield  {journal} {\bibinfo  {journal}
  {Astrophys. J. Suppl.}\ }\textbf {\bibinfo {volume} {194}},\ \bibinfo {pages}
  {39} (\bibinfo {year} {2011})}\BibitemShut {NoStop}%
\bibitem [{\citenamefont {Ouyed}\ \emph {et~al.}(2012)\citenamefont {Ouyed},
  \citenamefont {Kostka}, \citenamefont {Koning}, \citenamefont {Leahy},\ and\
  \citenamefont {Steffen}}]{2012MNRAS.423.1652O}%
  \BibitemOpen
  \bibfield  {author} {\bibinfo {author} {\bibfnamefont {R.}~\bibnamefont
  {Ouyed}}, \bibinfo {author} {\bibfnamefont {M.}~\bibnamefont {Kostka}},
  \bibinfo {author} {\bibfnamefont {N.}~\bibnamefont {Koning}}, \bibinfo
  {author} {\bibfnamefont {D.}~\bibnamefont {Leahy}}, \ and\ \bibinfo {author}
  {\bibfnamefont {W.}~\bibnamefont {Steffen}},\ }\href {\doibase
  10.1111/j.1365-2966.2012.20986.x} {\bibfield  {journal} {\bibinfo  {journal}
  {Mon. Not. Roy. Astron. Soc.}\ }\textbf {\bibinfo {volume} {423}},\ \bibinfo
  {pages} {1652} (\bibinfo {year} {2012})}\BibitemShut {NoStop}%
\bibitem [{\citenamefont {Fischer}\ \emph {et~al.}(2017)\citenamefont
  {Fischer}, \citenamefont {Bastian}, \citenamefont {Wu}, \citenamefont
  {Typel}, \citenamefont {Klahn},\ and\ \citenamefont
  {Blaschke}}]{Fischer:2017lag}%
  \BibitemOpen
  \bibfield  {author} {\bibinfo {author} {\bibfnamefont {T.}~\bibnamefont
  {Fischer}}, \bibinfo {author} {\bibfnamefont {N.-U.~F.}\ \bibnamefont
  {Bastian}}, \bibinfo {author} {\bibfnamefont {M.-R.}\ \bibnamefont {Wu}},
  \bibinfo {author} {\bibfnamefont {S.}~\bibnamefont {Typel}}, \bibinfo
  {author} {\bibfnamefont {T.}~\bibnamefont {Klahn}}, \ and\ \bibinfo {author}
  {\bibfnamefont {D.~B.}\ \bibnamefont {Blaschke}},\ }\href@noop {} {\
  (\bibinfo {year} {2017})},\ \Eprint {http://arxiv.org/abs/1712.08788}
  {arXiv:1712.08788 [astro-ph.HE]} \BibitemShut {NoStop}%
\bibitem [{\citenamefont {Drago}\ and\ \citenamefont
  {Tambini}(1999)}]{Drago:1997tn}%
  \BibitemOpen
  \bibfield  {author} {\bibinfo {author} {\bibfnamefont {A.}~\bibnamefont
  {Drago}}\ and\ \bibinfo {author} {\bibfnamefont {U.}~\bibnamefont
  {Tambini}},\ }\href {\doibase 10.1088/0954-3899/25/5/302} {\bibfield
  {journal} {\bibinfo  {journal} {J. Phys.}\ }\textbf {\bibinfo {volume}
  {G25}},\ \bibinfo {pages} {971} (\bibinfo {year} {1999})}\BibitemShut
  {NoStop}%
\bibitem [{\citenamefont {Lugones}\ and\ \citenamefont
  {Grunfeld}(2011)}]{Lugones:2011xv}%
  \BibitemOpen
  \bibfield  {author} {\bibinfo {author} {\bibfnamefont {G.}~\bibnamefont
  {Lugones}}\ and\ \bibinfo {author} {\bibfnamefont {A.~G.}\ \bibnamefont
  {Grunfeld}},\ }\href {\doibase 10.1103/PhysRevD.84.085003} {\bibfield
  {journal} {\bibinfo  {journal} {Phys. Rev.}\ }\textbf {\bibinfo {volume}
  {D84}},\ \bibinfo {pages} {085003} (\bibinfo {year} {2011})}\BibitemShut
  {NoStop}%
\bibitem [{\citenamefont {Bombaci}\ \emph {et~al.}(2011)\citenamefont
  {Bombaci}, \citenamefont {Logoteta}, \citenamefont {Providencia},\ and\
  \citenamefont {Vidana}}]{Bombaci:2011mx}%
  \BibitemOpen
  \bibfield  {author} {\bibinfo {author} {\bibfnamefont {I.}~\bibnamefont
  {Bombaci}}, \bibinfo {author} {\bibfnamefont {D.}~\bibnamefont {Logoteta}},
  \bibinfo {author} {\bibfnamefont {C.}~\bibnamefont {Providencia}}, \ and\
  \bibinfo {author} {\bibfnamefont {I.}~\bibnamefont {Vidana}},\ }\href
  {\doibase 10.1051/0004-6361/201015783} {\bibfield  {journal} {\bibinfo
  {journal} {Astron. Astrophys.}\ }\textbf {\bibinfo {volume} {528}},\ \bibinfo
  {pages} {A71} (\bibinfo {year} {2011})}\BibitemShut {NoStop}%
\bibitem [{\citenamefont {Grunfeld}\ and\ \citenamefont
  {Lugones}(2018)}]{Grunfeld:2018lxb}%
  \BibitemOpen
  \bibfield  {author} {\bibinfo {author} {\bibfnamefont {A.~G.}\ \bibnamefont
  {Grunfeld}}\ and\ \bibinfo {author} {\bibfnamefont {G.}~\bibnamefont
  {Lugones}},\ }in\ \href
  {https://inspirehep.net/record/1670155/files/1804.09898.pdf} {\emph {\bibinfo
  {booktitle} {{14th International Workshop on Hadron Physics (Hadron Physics
  2018) Florianopolis, Santa Catarina, Brazil, March 18-23, 2018}}}}\ (\bibinfo
  {year} {2018})\BibitemShut {NoStop}%
\bibitem [{\citenamefont {Yasutake}\ \emph {et~al.}(2014)\citenamefont
  {Yasutake}, \citenamefont {Lastowiecki}, \citenamefont {Benic}, \citenamefont
  {Blaschke}, \citenamefont {Maruyama},\ and\ \citenamefont
  {Tatsumi}}]{Yasutake:2014oxa}%
  \BibitemOpen
  \bibfield  {author} {\bibinfo {author} {\bibfnamefont {N.}~\bibnamefont
  {Yasutake}}, \bibinfo {author} {\bibfnamefont {R.}~\bibnamefont
  {Lastowiecki}}, \bibinfo {author} {\bibfnamefont {S.}~\bibnamefont {Benic}},
  \bibinfo {author} {\bibfnamefont {D.}~\bibnamefont {Blaschke}}, \bibinfo
  {author} {\bibfnamefont {T.}~\bibnamefont {Maruyama}}, \ and\ \bibinfo
  {author} {\bibfnamefont {T.}~\bibnamefont {Tatsumi}},\ }\href {\doibase
  10.1103/PhysRevC.89.065803} {\bibfield  {journal} {\bibinfo  {journal} {Phys.
  Rev.}\ }\textbf {\bibinfo {volume} {C89}},\ \bibinfo {pages} {065803}
  (\bibinfo {year} {2014})}\BibitemShut {NoStop}%
\bibitem [{\citenamefont {Ayriyan}\ and\ \citenamefont
  {Grigorian}(2018)}]{Ayriyan:2017tvl}%
  \BibitemOpen
  \bibfield  {author} {\bibinfo {author} {\bibfnamefont {A.}~\bibnamefont
  {Ayriyan}}\ and\ \bibinfo {author} {\bibfnamefont {H.}~\bibnamefont
  {Grigorian}},\ }\bibfield  {booktitle} {\emph {\bibinfo {booktitle}
  {{Proceedings, International Conference "Mathematical Modeling and
  Computational Physics, 2017" (MMCP2017): Dubna, Russia, July 3-7, 2017}}},\
  }\href {\doibase 10.1051/epjconf/201817303003} {\bibfield  {journal}
  {\bibinfo  {journal} {EPJ Web Conf.}\ }\textbf {\bibinfo {volume} {173}},\
  \bibinfo {pages} {03003} (\bibinfo {year} {2018})}\BibitemShut {NoStop}%
\bibitem [{\citenamefont {Reid}\ and\ \citenamefont {Sherwood}()}]{Reid:1966}%
  \BibitemOpen
  \bibfield  {author} {\bibinfo {author} {\bibfnamefont {R.~C.}\ \bibnamefont
  {Reid}}\ and\ \bibinfo {author} {\bibfnamefont {T.~K.}\ \bibnamefont
  {Sherwood}},\ }\href@noop {} {\emph {\bibinfo {title} {The properties of
  gases and liquids}}}\ (\bibinfo  {publisher} {McGraw-Hill, New York,
  1966})\BibitemShut {NoStop}%
\bibitem [{\citenamefont {Landau}\ and\ \citenamefont
  {Lifshitz}()}]{Landau:1969}%
  \BibitemOpen
  \bibfield  {author} {\bibinfo {author} {\bibfnamefont {L.~D.}\ \bibnamefont
  {Landau}}\ and\ \bibinfo {author} {\bibfnamefont {E.~M.}\ \bibnamefont
  {Lifshitz}},\ }\href@noop {} {\emph {\bibinfo {title} {Statistical Physics.
  Pt.1}}}\ (\bibinfo  {publisher} {Oxford: Pergamom Press, and Reading:
  Addison-Wesley, 1969})\BibitemShut {NoStop}%
\bibitem [{\citenamefont {Greiner}\ \emph {et~al.}(1987)\citenamefont
  {Greiner}, \citenamefont {Koch},\ and\ \citenamefont
  {Stoecker}}]{Greiner:1987tg}%
  \BibitemOpen
  \bibfield  {author} {\bibinfo {author} {\bibfnamefont {C.}~\bibnamefont
  {Greiner}}, \bibinfo {author} {\bibfnamefont {P.}~\bibnamefont {Koch}}, \
  and\ \bibinfo {author} {\bibfnamefont {H.}~\bibnamefont {Stoecker}},\ }\href
  {\doibase 10.1103/PhysRevLett.58.1825} {\bibfield  {journal} {\bibinfo
  {journal} {Phys. Rev. Lett.}\ }\textbf {\bibinfo {volume} {58}},\ \bibinfo
  {pages} {1825} (\bibinfo {year} {1987})}\BibitemShut {NoStop}%
\bibitem [{\citenamefont {Clarke}\ \emph {et~al.}(1994)\citenamefont {Clarke},
  \citenamefont {Hastie}, \citenamefont {Kihlborg}, \citenamefont {Metselaar},\
  and\ \citenamefont {Thackeray}}]{Clarke:1994}%
  \BibitemOpen
  \bibfield  {author} {\bibinfo {author} {\bibfnamefont {J.~B.}\ \bibnamefont
  {Clarke}}, \bibinfo {author} {\bibfnamefont {J.~W.}\ \bibnamefont {Hastie}},
  \bibinfo {author} {\bibfnamefont {L.~H.~E.}\ \bibnamefont {Kihlborg}},
  \bibinfo {author} {\bibfnamefont {R.}~\bibnamefont {Metselaar}}, \ and\
  \bibinfo {author} {\bibfnamefont {M.~M.}\ \bibnamefont {Thackeray}},\
  }\href@noop {} {\bibfield  {journal} {\bibinfo  {journal} {Pure and Appl.
  Chem.}\ }\textbf {\bibinfo {volume} {66}},\ \bibinfo {pages} {577} (\bibinfo
  {year} {1994})}\BibitemShut {NoStop}%
\bibitem [{\citenamefont {Iosilevskiy}\ \emph {et~al.}(2003)\citenamefont
  {Iosilevskiy}, \citenamefont {Gryaznov}, \citenamefont {Yakub}, \citenamefont
  {Ronchi},\ and\ \citenamefont {Fortov}}]{Iosilevskiy:2003}%
  \BibitemOpen
  \bibfield  {author} {\bibinfo {author} {\bibfnamefont {I.}~\bibnamefont
  {Iosilevskiy}}, \bibinfo {author} {\bibfnamefont {V.}~\bibnamefont
  {Gryaznov}}, \bibinfo {author} {\bibfnamefont {E.}~\bibnamefont {Yakub}},
  \bibinfo {author} {\bibfnamefont {C.}~\bibnamefont {Ronchi}}, \ and\ \bibinfo
  {author} {\bibfnamefont {V.}~\bibnamefont {Fortov}},\ }\href@noop {}
  {\bibfield  {journal} {\bibinfo  {journal} {Contributions to Plasma Physics}\
  }\textbf {\bibinfo {volume} {43}},\ \bibinfo {pages} {316} (\bibinfo {year}
  {2003})}\BibitemShut {NoStop}%
\bibitem [{\citenamefont {Iosilevskiy}(2010)}]{Iosilevskiy:2010qr}%
  \BibitemOpen
  \bibfield  {author} {\bibinfo {author} {\bibfnamefont {I.}~\bibnamefont
  {Iosilevskiy}},\ }\bibfield  {booktitle} {\emph {\bibinfo {booktitle} {{Three
  days of strong interactions. Proceedings, EMMI Workshop and 26th Max Born
  Symposium, Wroclaw, Poland, July 9-11, 2009}}},\ }\href@noop {} {\bibfield
  {journal} {\bibinfo  {journal} {Acta Phys. Polon. Supp.}\ }\textbf {\bibinfo
  {volume} {3}},\ \bibinfo {pages} {589} (\bibinfo {year} {2010})}\BibitemShut
  {NoStop}%
\bibitem [{\citenamefont {Andersen}\ and\ \citenamefont
  {Strickland}(2002)}]{Andersen:2002jz}%
  \BibitemOpen
  \bibfield  {author} {\bibinfo {author} {\bibfnamefont {J.~O.}\ \bibnamefont
  {Andersen}}\ and\ \bibinfo {author} {\bibfnamefont {M.}~\bibnamefont
  {Strickland}},\ }\href {\doibase 10.1103/PhysRevD.66.105001} {\bibfield
  {journal} {\bibinfo  {journal} {Phys. Rev.}\ }\textbf {\bibinfo {volume}
  {D66}},\ \bibinfo {pages} {105001} (\bibinfo {year} {2002})}\BibitemShut
  {NoStop}%
\bibitem [{\citenamefont {Fraga}\ \emph {et~al.}(2014)\citenamefont {Fraga},
  \citenamefont {Kurkela},\ and\ \citenamefont {Vuorinen}}]{Fraga:2013qra}%
  \BibitemOpen
  \bibfield  {author} {\bibinfo {author} {\bibfnamefont {E.~S.}\ \bibnamefont
  {Fraga}}, \bibinfo {author} {\bibfnamefont {A.}~\bibnamefont {Kurkela}}, \
  and\ \bibinfo {author} {\bibfnamefont {A.}~\bibnamefont {Vuorinen}},\ }\href
  {\doibase 10.1088/2041-8205/781/2/L25} {\bibfield  {journal} {\bibinfo
  {journal} {Astrophys. J.}\ }\textbf {\bibinfo {volume} {781}},\ \bibinfo
  {pages} {L25} (\bibinfo {year} {2014})}\BibitemShut {NoStop}%
\bibitem [{\citenamefont {Kurkela}\ and\ \citenamefont
  {Vuorinen}(2016)}]{Kurkela:2016was}%
  \BibitemOpen
  \bibfield  {author} {\bibinfo {author} {\bibfnamefont {A.}~\bibnamefont
  {Kurkela}}\ and\ \bibinfo {author} {\bibfnamefont {A.}~\bibnamefont
  {Vuorinen}},\ }\href {\doibase 10.1103/PhysRevLett.117.042501} {\bibfield
  {journal} {\bibinfo  {journal} {Phys. Rev. Lett.}\ }\textbf {\bibinfo
  {volume} {117}},\ \bibinfo {pages} {042501} (\bibinfo {year}
  {2016})}\BibitemShut {NoStop}%
\bibitem [{\citenamefont {Muroya}\ \emph {et~al.}(2003)\citenamefont {Muroya},
  \citenamefont {Nakamura}, \citenamefont {Nonaka},\ and\ \citenamefont
  {Takaishi}}]{Muroya:2003qs}%
  \BibitemOpen
  \bibfield  {author} {\bibinfo {author} {\bibfnamefont {S.}~\bibnamefont
  {Muroya}}, \bibinfo {author} {\bibfnamefont {A.}~\bibnamefont {Nakamura}},
  \bibinfo {author} {\bibfnamefont {C.}~\bibnamefont {Nonaka}}, \ and\ \bibinfo
  {author} {\bibfnamefont {T.}~\bibnamefont {Takaishi}},\ }\href {\doibase
  10.1143/PTP.110.615} {\bibfield  {journal} {\bibinfo  {journal} {Prog. Theor.
  Phys.}\ }\textbf {\bibinfo {volume} {110}},\ \bibinfo {pages} {615} (\bibinfo
  {year} {2003})}\BibitemShut {NoStop}%
\bibitem [{\citenamefont {de~Forcrand}(2009)}]{deForcrand:2010ys}%
  \BibitemOpen
  \bibfield  {author} {\bibinfo {author} {\bibfnamefont {P.}~\bibnamefont
  {de~Forcrand}},\ }\bibfield  {booktitle} {\emph {\bibinfo {booktitle}
  {{Proceedings, 27th International Symposium on Lattice field theory (Lattice
  2009): Beijing, P.R. China, July 26-31, 2009}}},\ }\href@noop {} {\bibfield
  {journal} {\bibinfo  {journal} {PoS}\ }\textbf {\bibinfo {volume}
  {LAT2009}},\ \bibinfo {pages} {010} (\bibinfo {year} {2009})}\BibitemShut
  {NoStop}%
\bibitem [{\citenamefont {Dexheimer}\ and\ \citenamefont
  {Schramm}(2008)}]{Dexheimer:2008ax}%
  \BibitemOpen
  \bibfield  {author} {\bibinfo {author} {\bibfnamefont {V.}~\bibnamefont
  {Dexheimer}}\ and\ \bibinfo {author} {\bibfnamefont {S.}~\bibnamefont
  {Schramm}},\ }\href {\doibase 10.1086/589735} {\bibfield  {journal} {\bibinfo
   {journal} {Astrophys. J.}\ }\textbf {\bibinfo {volume} {683}},\ \bibinfo
  {pages} {943} (\bibinfo {year} {2008})}\BibitemShut {NoStop}%
\bibitem [{\citenamefont {Franzon}\ \emph {et~al.}(2016)\citenamefont
  {Franzon}, \citenamefont {Dexheimer},\ and\ \citenamefont
  {Schramm}}]{Franzon:2016iai}%
  \BibitemOpen
  \bibfield  {author} {\bibinfo {author} {\bibfnamefont {B.}~\bibnamefont
  {Franzon}}, \bibinfo {author} {\bibfnamefont {V.}~\bibnamefont {Dexheimer}},
  \ and\ \bibinfo {author} {\bibfnamefont {S.}~\bibnamefont {Schramm}},\ }\href
  {\doibase 10.1103/PhysRevD.94.044018} {\bibfield  {journal} {\bibinfo
  {journal} {Phys. Rev.}\ }\textbf {\bibinfo {volume} {D94}},\ \bibinfo {pages}
  {044018} (\bibinfo {year} {2016})}\BibitemShut {NoStop}%
\bibitem [{\citenamefont {Dexheimer}\ and\ \citenamefont
  {Schramm}(2010)}]{Dexheimer:2009hi}%
  \BibitemOpen
  \bibfield  {author} {\bibinfo {author} {\bibfnamefont {V.~A.}\ \bibnamefont
  {Dexheimer}}\ and\ \bibinfo {author} {\bibfnamefont {S.}~\bibnamefont
  {Schramm}},\ }\href {\doibase 10.1103/PhysRevC.81.045201} {\bibfield
  {journal} {\bibinfo  {journal} {Phys. Rev.}\ }\textbf {\bibinfo {volume}
  {C81}},\ \bibinfo {pages} {045201} (\bibinfo {year} {2010})}\BibitemShut
  {NoStop}%
\bibitem [{\citenamefont {Dexheimer}\ \emph {et~al.}(2013)\citenamefont
  {Dexheimer}, \citenamefont {Negreiros}, \citenamefont {Schramm},\ and\
  \citenamefont {Hempel}}]{Dexheimer:2012qk}%
  \BibitemOpen
  \bibfield  {author} {\bibinfo {author} {\bibfnamefont {V.}~\bibnamefont
  {Dexheimer}}, \bibinfo {author} {\bibfnamefont {R.}~\bibnamefont
  {Negreiros}}, \bibinfo {author} {\bibfnamefont {S.}~\bibnamefont {Schramm}},
  \ and\ \bibinfo {author} {\bibfnamefont {M.}~\bibnamefont {Hempel}},\
  }\bibfield  {booktitle} {\emph {\bibinfo {booktitle} {{Proceedings, 12th
  International Workshop on Hadron Physics: Bento Goncalves, Rio Grande do Sul,
  Brazil, April 22-27, 2012}}},\ }\href {\doibase 10.1063/1.4795968} {\bibfield
   {journal} {\bibinfo  {journal} {AIP Conf. Proc.}\ }\textbf {\bibinfo
  {volume} {1520}},\ \bibinfo {pages} {264} (\bibinfo {year}
  {2013})}\BibitemShut {NoStop}%
\bibitem [{\citenamefont {Dexheimer}\ \emph {et~al.}(2017)\citenamefont
  {Dexheimer}, \citenamefont {Hempel}, \citenamefont {Iosilevskiy},\ and\
  \citenamefont {Schramm}}]{Dexheimer:2017ecc}%
  \BibitemOpen
  \bibfield  {author} {\bibinfo {author} {\bibfnamefont {V.}~\bibnamefont
  {Dexheimer}}, \bibinfo {author} {\bibfnamefont {M.}~\bibnamefont {Hempel}},
  \bibinfo {author} {\bibfnamefont {I.}~\bibnamefont {Iosilevskiy}}, \ and\
  \bibinfo {author} {\bibfnamefont {S.}~\bibnamefont {Schramm}},\ }\bibfield
  {booktitle} {\emph {\bibinfo {booktitle} {{Proceedings, 26th International
  Conference on Ultra-relativistic Nucleus-Nucleus Collisions (Quark Matter
  2017): Chicago, Illinois, USA, February 5-11, 2017}}},\ }\href {\doibase
  10.1016/j.nuclphysa.2017.04.031} {\bibfield  {journal} {\bibinfo  {journal}
  {Nucl. Phys.}\ }\textbf {\bibinfo {volume} {A967}},\ \bibinfo {pages} {780}
  (\bibinfo {year} {2017})}\BibitemShut {NoStop}%
\bibitem [{\citenamefont {Papazoglou}\ \emph {et~al.}(1999)\citenamefont
  {Papazoglou}, \citenamefont {Zschiesche}, \citenamefont {Schramm},
  \citenamefont {Schaffner-Bielich}, \citenamefont {Stoecker},\ and\
  \citenamefont {Greiner}}]{Papazoglou:1998vr}%
  \BibitemOpen
  \bibfield  {author} {\bibinfo {author} {\bibfnamefont {P.}~\bibnamefont
  {Papazoglou}}, \bibinfo {author} {\bibfnamefont {D.}~\bibnamefont
  {Zschiesche}}, \bibinfo {author} {\bibfnamefont {S.}~\bibnamefont {Schramm}},
  \bibinfo {author} {\bibfnamefont {J.}~\bibnamefont {Schaffner-Bielich}},
  \bibinfo {author} {\bibfnamefont {H.}~\bibnamefont {Stoecker}}, \ and\
  \bibinfo {author} {\bibfnamefont {W.}~\bibnamefont {Greiner}},\ }\href
  {\doibase 10.1103/PhysRevC.59.411} {\bibfield  {journal} {\bibinfo  {journal}
  {Phys. Rev.}\ }\textbf {\bibinfo {volume} {C59}},\ \bibinfo {pages} {411}
  (\bibinfo {year} {1999})}\BibitemShut {NoStop}%
\bibitem [{\citenamefont {Ratti}\ \emph
  {et~al.}(2006{\natexlab{a}})\citenamefont {Ratti}, \citenamefont {Thaler},\
  and\ \citenamefont {Weise}}]{Ratti:2006ka}%
  \BibitemOpen
  \bibfield  {author} {\bibinfo {author} {\bibfnamefont {C.}~\bibnamefont
  {Ratti}}, \bibinfo {author} {\bibfnamefont {M.~A.}\ \bibnamefont {Thaler}}, \
  and\ \bibinfo {author} {\bibfnamefont {W.}~\bibnamefont {Weise}},\ }\bibfield
   {booktitle} {\emph {\bibinfo {booktitle} {{Proceedings, 18th International
  Conference on Ultra-Relativistic Nucleus-Nucleus Collisions (Quark Matter
  2005): Budapest, Hungary, August 4-9, 2005}}},\ }\href@noop {} {\bibfield
  {journal} {\bibinfo  {journal} {Rom. Rep. Phys.}\ }\textbf {\bibinfo {volume}
  {58}},\ \bibinfo {pages} {13} (\bibinfo {year}
  {2006}{\natexlab{a}})}\BibitemShut {NoStop}%
\bibitem [{\citenamefont {Roessner}\ \emph {et~al.}(2007)\citenamefont
  {Roessner}, \citenamefont {Ratti},\ and\ \citenamefont
  {Weise}}]{Roessner:2006xn}%
  \BibitemOpen
  \bibfield  {author} {\bibinfo {author} {\bibfnamefont {S.}~\bibnamefont
  {Roessner}}, \bibinfo {author} {\bibfnamefont {C.}~\bibnamefont {Ratti}}, \
  and\ \bibinfo {author} {\bibfnamefont {W.}~\bibnamefont {Weise}},\ }\href
  {\doibase 10.1103/PhysRevD.75.034007} {\bibfield  {journal} {\bibinfo
  {journal} {Phys. Rev.}\ }\textbf {\bibinfo {volume} {D75}},\ \bibinfo {pages}
  {034007} (\bibinfo {year} {2007})}\BibitemShut {NoStop}%
\bibitem [{\citenamefont {Ratti}\ \emph
  {et~al.}(2006{\natexlab{b}})\citenamefont {Ratti}, \citenamefont {Thaler},\
  and\ \citenamefont {Weise}}]{Ratti:2005jh}%
  \BibitemOpen
  \bibfield  {author} {\bibinfo {author} {\bibfnamefont {C.}~\bibnamefont
  {Ratti}}, \bibinfo {author} {\bibfnamefont {M.~A.}\ \bibnamefont {Thaler}}, \
  and\ \bibinfo {author} {\bibfnamefont {W.}~\bibnamefont {Weise}},\ }\href
  {\doibase 10.1103/PhysRevD.73.014019} {\bibfield  {journal} {\bibinfo
  {journal} {Phys. Rev.}\ }\textbf {\bibinfo {volume} {D73}},\ \bibinfo {pages}
  {014019} (\bibinfo {year} {2006}{\natexlab{b}})}\BibitemShut {NoStop}%
\bibitem [{\citenamefont {Fodor}\ and\ \citenamefont
  {Katz}(2004)}]{Fodor:2004nz}%
  \BibitemOpen
  \bibfield  {author} {\bibinfo {author} {\bibfnamefont {Z.}~\bibnamefont
  {Fodor}}\ and\ \bibinfo {author} {\bibfnamefont {S.~D.}\ \bibnamefont
  {Katz}},\ }\href {\doibase 10.1088/1126-6708/2004/04/050} {\bibfield
  {journal} {\bibinfo  {journal} {JHEP}\ }\textbf {\bibinfo {volume} {04}},\
  \bibinfo {pages} {050} (\bibinfo {year} {2004})}\BibitemShut {NoStop}%
\bibitem [{\citenamefont {Aoki}\ \emph {et~al.}(2006)\citenamefont {Aoki},
  \citenamefont {Endrodi}, \citenamefont {Fodor}, \citenamefont {Katz},\ and\
  \citenamefont {Szabo}}]{Aoki:2006we}%
  \BibitemOpen
  \bibfield  {author} {\bibinfo {author} {\bibfnamefont {Y.}~\bibnamefont
  {Aoki}}, \bibinfo {author} {\bibfnamefont {G.}~\bibnamefont {Endrodi}},
  \bibinfo {author} {\bibfnamefont {Z.}~\bibnamefont {Fodor}}, \bibinfo
  {author} {\bibfnamefont {S.~D.}\ \bibnamefont {Katz}}, \ and\ \bibinfo
  {author} {\bibfnamefont {K.~K.}\ \bibnamefont {Szabo}},\ }\href {\doibase
  10.1038/nature05120} {\bibfield  {journal} {\bibinfo  {journal} {Nature}\
  }\textbf {\bibinfo {volume} {443}},\ \bibinfo {pages} {675} (\bibinfo {year}
  {2006})}\BibitemShut {NoStop}%
\bibitem [{\citenamefont {Fischer}\ \emph {et~al.}(2010)\citenamefont
  {Fischer}, \citenamefont {Whitehouse}, \citenamefont {Mezzacappa},
  \citenamefont {Thielemann},\ and\ \citenamefont
  {Liebendorfer}}]{Fischer:2009af}%
  \BibitemOpen
  \bibfield  {author} {\bibinfo {author} {\bibfnamefont {T.}~\bibnamefont
  {Fischer}}, \bibinfo {author} {\bibfnamefont {S.~C.}\ \bibnamefont
  {Whitehouse}}, \bibinfo {author} {\bibfnamefont {A.}~\bibnamefont
  {Mezzacappa}}, \bibinfo {author} {\bibfnamefont {F.~K.}\ \bibnamefont
  {Thielemann}}, \ and\ \bibinfo {author} {\bibfnamefont {M.}~\bibnamefont
  {Liebendorfer}},\ }\href {\doibase 10.1051/0004-6361/200913106} {\bibfield
  {journal} {\bibinfo  {journal} {Astron. Astrophys.}\ }\textbf {\bibinfo
  {volume} {517}},\ \bibinfo {pages} {A80} (\bibinfo {year}
  {2010})}\BibitemShut {NoStop}%
\bibitem [{\citenamefont {Hudepohl}\ \emph {et~al.}(2010)\citenamefont
  {Hudepohl}, \citenamefont {Muller}, \citenamefont {Janka}, \citenamefont
  {Marek},\ and\ \citenamefont {Raffelt}}]{Huedepohl:2009wh}%
  \BibitemOpen
  \bibfield  {author} {\bibinfo {author} {\bibfnamefont {L.}~\bibnamefont
  {Hudepohl}}, \bibinfo {author} {\bibfnamefont {B.}~\bibnamefont {Muller}},
  \bibinfo {author} {\bibfnamefont {H.~T.}\ \bibnamefont {Janka}}, \bibinfo
  {author} {\bibfnamefont {A.}~\bibnamefont {Marek}}, \ and\ \bibinfo {author}
  {\bibfnamefont {G.~G.}\ \bibnamefont {Raffelt}},\ }\href {\doibase
  10.1103/PhysRevLett.104.251101, 10.1103/PhysRevLett.105.249901} {\bibfield
  {journal} {\bibinfo  {journal} {Phys. Rev. Lett.}\ }\textbf {\bibinfo
  {volume} {104}},\ \bibinfo {pages} {251101} (\bibinfo {year} {2010})},\
  \bibinfo {note} {[Erratum: Phys. Rev. Lett.105,249901(2010)]}\BibitemShut
  {NoStop}%
\bibitem [{\citenamefont {Hempel}\ \emph {et~al.}(2009)\citenamefont {Hempel},
  \citenamefont {Pagliara},\ and\ \citenamefont
  {Schaffner-Bielich}}]{Hempel:2009vp}%
  \BibitemOpen
  \bibfield  {author} {\bibinfo {author} {\bibfnamefont {M.}~\bibnamefont
  {Hempel}}, \bibinfo {author} {\bibfnamefont {G.}~\bibnamefont {Pagliara}}, \
  and\ \bibinfo {author} {\bibfnamefont {J.}~\bibnamefont
  {Schaffner-Bielich}},\ }\href {\doibase 10.1103/PhysRevD.80.125014}
  {\bibfield  {journal} {\bibinfo  {journal} {Phys. Rev.}\ }\textbf {\bibinfo
  {volume} {D80}},\ \bibinfo {pages} {125014} (\bibinfo {year}
  {2009})}\BibitemShut {NoStop}%
\bibitem [{\citenamefont {Garcia}\ and\ \citenamefont
  {Pinto}(2013)}]{Garcia:2013eaa}%
  \BibitemOpen
  \bibfield  {author} {\bibinfo {author} {\bibfnamefont {A.~F.}\ \bibnamefont
  {Garcia}}\ and\ \bibinfo {author} {\bibfnamefont {M.~B.}\ \bibnamefont
  {Pinto}},\ }\href {\doibase 10.1103/PhysRevC.88.025207} {\bibfield  {journal}
  {\bibinfo  {journal} {Phys. Rev.}\ }\textbf {\bibinfo {volume} {C88}},\
  \bibinfo {pages} {025207} (\bibinfo {year} {2013})}\BibitemShut {NoStop}%
\bibitem [{\citenamefont {Lugones}\ \emph {et~al.}(2013)\citenamefont
  {Lugones}, \citenamefont {Grunfeld},\ and\ \citenamefont
  {Al~Ajmi}}]{Lugones:2013ema}%
  \BibitemOpen
  \bibfield  {author} {\bibinfo {author} {\bibfnamefont {G.}~\bibnamefont
  {Lugones}}, \bibinfo {author} {\bibfnamefont {A.~G.}\ \bibnamefont
  {Grunfeld}}, \ and\ \bibinfo {author} {\bibfnamefont {M.}~\bibnamefont
  {Al~Ajmi}},\ }\href {\doibase 10.1103/PhysRevC.88.045803} {\bibfield
  {journal} {\bibinfo  {journal} {Phys. Rev.}\ }\textbf {\bibinfo {volume}
  {C88}},\ \bibinfo {pages} {045803} (\bibinfo {year} {2013})}\BibitemShut
  {NoStop}%
\bibitem [{\citenamefont {Mariani}\ \emph {et~al.}(2017)\citenamefont
  {Mariani}, \citenamefont {Orsaria},\ and\ \citenamefont
  {Vucetich}}]{Mariani:2016pcx}%
  \BibitemOpen
  \bibfield  {author} {\bibinfo {author} {\bibfnamefont {M.}~\bibnamefont
  {Mariani}}, \bibinfo {author} {\bibfnamefont {M.}~\bibnamefont {Orsaria}}, \
  and\ \bibinfo {author} {\bibfnamefont {H.}~\bibnamefont {Vucetich}},\ }\href
  {\doibase 10.1051/0004-6361/201629315} {\bibfield  {journal} {\bibinfo
  {journal} {Astron. Astrophys.}\ }\textbf {\bibinfo {volume} {601}},\ \bibinfo
  {pages} {A21} (\bibinfo {year} {2017})}\BibitemShut {NoStop}%
\bibitem [{\citenamefont {Nicotra}\ \emph {et~al.}(2006)\citenamefont
  {Nicotra}, \citenamefont {Baldo}, \citenamefont {Burgio},\ and\ \citenamefont
  {Schulze}}]{Nicotra:2005fj}%
  \BibitemOpen
  \bibfield  {author} {\bibinfo {author} {\bibfnamefont {O.~E.}\ \bibnamefont
  {Nicotra}}, \bibinfo {author} {\bibfnamefont {M.}~\bibnamefont {Baldo}},
  \bibinfo {author} {\bibfnamefont {G.~F.}\ \bibnamefont {Burgio}}, \ and\
  \bibinfo {author} {\bibfnamefont {H.~J.}\ \bibnamefont {Schulze}},\ }\href
  {\doibase 10.1051/0004-6361:20053670} {\bibfield  {journal} {\bibinfo
  {journal} {Astron. Astrophys.}\ }\textbf {\bibinfo {volume} {451}},\ \bibinfo
  {pages} {213} (\bibinfo {year} {2006})}\BibitemShut {NoStop}%
\bibitem [{\citenamefont {Burgio}\ \emph {et~al.}(2007)\citenamefont {Burgio},
  \citenamefont {Baldo}, \citenamefont {Nicotra},\ and\ \citenamefont
  {Schulze}}]{Burgio:2006ed}%
  \BibitemOpen
  \bibfield  {author} {\bibinfo {author} {\bibfnamefont {G.~F.}\ \bibnamefont
  {Burgio}}, \bibinfo {author} {\bibfnamefont {M.}~\bibnamefont {Baldo}},
  \bibinfo {author} {\bibfnamefont {O.~E.}\ \bibnamefont {Nicotra}}, \ and\
  \bibinfo {author} {\bibfnamefont {H.~J.}\ \bibnamefont {Schulze}},\
  }\bibfield  {booktitle} {\emph {\bibinfo {booktitle} {{Conference on Isolated
  Neutron Stars: From the Interior to the Surface London, England, April 24-28,
  2006}}},\ }\href {\doibase 10.1007/s10509-007-9360-8} {\bibfield  {journal}
  {\bibinfo  {journal} {Astrophys. Space Sci.}\ }\textbf {\bibinfo {volume}
  {308}},\ \bibinfo {pages} {387} (\bibinfo {year} {2007})}\BibitemShut
  {NoStop}%
\bibitem [{\citenamefont {Braaten}\ and\ \citenamefont
  {Nieto}(1995)}]{Braaten:1995cm}%
  \BibitemOpen
  \bibfield  {author} {\bibinfo {author} {\bibfnamefont {E.}~\bibnamefont
  {Braaten}}\ and\ \bibinfo {author} {\bibfnamefont {A.}~\bibnamefont
  {Nieto}},\ }\href {\doibase 10.1103/PhysRevD.51.6990} {\bibfield  {journal}
  {\bibinfo  {journal} {Phys. Rev.}\ }\textbf {\bibinfo {volume} {D51}},\
  \bibinfo {pages} {6990} (\bibinfo {year} {1995})}\BibitemShut {NoStop}%
\bibitem [{\citenamefont {Braaten}\ and\ \citenamefont
  {Pisarski}(1992)}]{Braaten:1991gm}%
  \BibitemOpen
  \bibfield  {author} {\bibinfo {author} {\bibfnamefont {E.}~\bibnamefont
  {Braaten}}\ and\ \bibinfo {author} {\bibfnamefont {R.~D.}\ \bibnamefont
  {Pisarski}},\ }\href {\doibase 10.1103/PhysRevD.45.R1827} {\bibfield
  {journal} {\bibinfo  {journal} {Phys. Rev.}\ }\textbf {\bibinfo {volume}
  {D45}},\ \bibinfo {pages} {R1827} (\bibinfo {year} {1992})}\BibitemShut
  {NoStop}%
\bibitem [{\citenamefont {Vuorinen}(2003)}]{Vuorinen:2003fs}%
  \BibitemOpen
  \bibfield  {author} {\bibinfo {author} {\bibfnamefont {A.}~\bibnamefont
  {Vuorinen}},\ }\href {\doibase 10.1103/PhysRevD.68.054017} {\bibfield
  {journal} {\bibinfo  {journal} {Phys. Rev.}\ }\textbf {\bibinfo {volume}
  {D68}},\ \bibinfo {pages} {054017} (\bibinfo {year} {2003})}\BibitemShut
  {NoStop}%
\bibitem [{\citenamefont {Kurkela}\ \emph {et~al.}(2014)\citenamefont
  {Kurkela}, \citenamefont {Fraga}, \citenamefont {Schaffner-Bielich},\ and\
  \citenamefont {Vuorinen}}]{Kurkela:2014vha}%
  \BibitemOpen
  \bibfield  {author} {\bibinfo {author} {\bibfnamefont {A.}~\bibnamefont
  {Kurkela}}, \bibinfo {author} {\bibfnamefont {E.~S.}\ \bibnamefont {Fraga}},
  \bibinfo {author} {\bibfnamefont {J.}~\bibnamefont {Schaffner-Bielich}}, \
  and\ \bibinfo {author} {\bibfnamefont {A.}~\bibnamefont {Vuorinen}},\ }\href
  {\doibase 10.1088/0004-637X/789/2/127} {\bibfield  {journal} {\bibinfo
  {journal} {Astrophys. J.}\ }\textbf {\bibinfo {volume} {789}},\ \bibinfo
  {pages} {127} (\bibinfo {year} {2014})}\BibitemShut {NoStop}%
\bibitem [{\citenamefont {Annala}\ \emph {et~al.}(2017)\citenamefont {Annala},
  \citenamefont {Gorda}, \citenamefont {Kurkela},\ and\ \citenamefont
  {Vuorinen}}]{Annala:2017llu}%
  \BibitemOpen
  \bibfield  {author} {\bibinfo {author} {\bibfnamefont {E.}~\bibnamefont
  {Annala}}, \bibinfo {author} {\bibfnamefont {T.}~\bibnamefont {Gorda}},
  \bibinfo {author} {\bibfnamefont {A.}~\bibnamefont {Kurkela}}, \ and\
  \bibinfo {author} {\bibfnamefont {A.}~\bibnamefont {Vuorinen}},\ }\href@noop
  {} {\  (\bibinfo {year} {2017})}\BibitemShut {NoStop}%
\bibitem [{\citenamefont {Jiménez}\ and\ \citenamefont
  {Fraga}(2018)}]{Jimenez:2017fax}%
  \BibitemOpen
  \bibfield  {author} {\bibinfo {author} {\bibfnamefont {J.~C.}\ \bibnamefont
  {Jiménez}}\ and\ \bibinfo {author} {\bibfnamefont {E.~S.}\ \bibnamefont
  {Fraga}},\ }\href {\doibase 10.1103/PhysRevD.97.094023} {\bibfield  {journal}
  {\bibinfo  {journal} {Phys. Rev.}\ }\textbf {\bibinfo {volume} {D97}},\
  \bibinfo {pages} {094023} (\bibinfo {year} {2018})}\BibitemShut {NoStop}%
\bibitem [{\citenamefont {Hempel}\ \emph {et~al.}(2017)\citenamefont {Hempel},
  \citenamefont {Heinimann}, \citenamefont {Yudin}, \citenamefont
  {Iosilevskiy}, \citenamefont {Liebendorfer},\ and\ \citenamefont
  {Friedrich-Karl}}]{Hempel:2017hsx}%
  \BibitemOpen
  \bibfield  {author} {\bibinfo {author} {\bibfnamefont {M.}~\bibnamefont
  {Hempel}}, \bibinfo {author} {\bibfnamefont {O.}~\bibnamefont {Heinimann}},
  \bibinfo {author} {\bibfnamefont {A.}~\bibnamefont {Yudin}}, \bibinfo
  {author} {\bibfnamefont {I.}~\bibnamefont {Iosilevskiy}}, \bibinfo {author}
  {\bibfnamefont {M.}~\bibnamefont {Liebendorfer}}, \ and\ \bibinfo {author}
  {\bibfnamefont {T.}~\bibnamefont {Friedrich-Karl}},\ }\href {\doibase
  10.1088/1742-6596/861/1/012023} {\bibfield  {journal} {\bibinfo  {journal}
  {J. Phys. Conf. Ser.}\ }\textbf {\bibinfo {volume} {861}},\ \bibinfo {pages}
  {012023} (\bibinfo {year} {2017})}\BibitemShut {NoStop}%
\bibitem [{\citenamefont {Bombaci}\ \emph {et~al.}(2009)\citenamefont
  {Bombaci}, \citenamefont {Logoteta}, \citenamefont {Panda}, \citenamefont
  {Providencia},\ and\ \citenamefont {Vidana}}]{Bombaci:2009jt}%
  \BibitemOpen
  \bibfield  {author} {\bibinfo {author} {\bibfnamefont {I.}~\bibnamefont
  {Bombaci}}, \bibinfo {author} {\bibfnamefont {D.}~\bibnamefont {Logoteta}},
  \bibinfo {author} {\bibfnamefont {P.~K.}\ \bibnamefont {Panda}}, \bibinfo
  {author} {\bibfnamefont {C.}~\bibnamefont {Providencia}}, \ and\ \bibinfo
  {author} {\bibfnamefont {I.}~\bibnamefont {Vidana}},\ }\href {\doibase
  10.1016/j.physletb.2009.09.039} {\bibfield  {journal} {\bibinfo  {journal}
  {Phys. Lett.}\ }\textbf {\bibinfo {volume} {B680}},\ \bibinfo {pages} {448}
  (\bibinfo {year} {2009})}\BibitemShut {NoStop}%
\bibitem [{\citenamefont {Yudin}\ \emph {et~al.}(2016)\citenamefont {Yudin},
  \citenamefont {Hempel}, \citenamefont {Nadyozhin},\ and\ \citenamefont
  {Razinkova}}]{Yudin:2015cva}%
  \BibitemOpen
  \bibfield  {author} {\bibinfo {author} {\bibfnamefont {A.~V.}\ \bibnamefont
  {Yudin}}, \bibinfo {author} {\bibfnamefont {M.}~\bibnamefont {Hempel}},
  \bibinfo {author} {\bibfnamefont {D.~K.}\ \bibnamefont {Nadyozhin}}, \ and\
  \bibinfo {author} {\bibfnamefont {T.~L.}\ \bibnamefont {Razinkova}},\ }\href
  {\doibase 10.1093/mnras/stv2614} {\bibfield  {journal} {\bibinfo  {journal}
  {Mon. Not. Roy. Astron. Soc.}\ }\textbf {\bibinfo {volume} {455}},\ \bibinfo
  {pages} {4325} (\bibinfo {year} {2016})}\BibitemShut {NoStop}%
\bibitem [{\citenamefont {Hempel}\ \emph {et~al.}(2016)\citenamefont {Hempel},
  \citenamefont {Heinimann}, \citenamefont {Yudin}, \citenamefont
  {Iosilevskiy}, \citenamefont {Liebendorfer},\ and\ \citenamefont
  {Thielemann}}]{Hempel:2015vlg}%
  \BibitemOpen
  \bibfield  {author} {\bibinfo {author} {\bibfnamefont {M.}~\bibnamefont
  {Hempel}}, \bibinfo {author} {\bibfnamefont {O.}~\bibnamefont {Heinimann}},
  \bibinfo {author} {\bibfnamefont {A.}~\bibnamefont {Yudin}}, \bibinfo
  {author} {\bibfnamefont {I.}~\bibnamefont {Iosilevskiy}}, \bibinfo {author}
  {\bibfnamefont {M.}~\bibnamefont {Liebendorfer}}, \ and\ \bibinfo {author}
  {\bibfnamefont {F.-K.}\ \bibnamefont {Thielemann}},\ }\href {\doibase
  10.1103/PhysRevD.94.103001} {\bibfield  {journal} {\bibinfo  {journal} {Phys.
  Rev.}\ }\textbf {\bibinfo {volume} {D94}},\ \bibinfo {pages} {103001}
  (\bibinfo {year} {2016})}\BibitemShut {NoStop}%
\bibitem [{\citenamefont {Fischer}\ \emph {et~al.}(2012)\citenamefont {Fischer}
  \emph {et~al.}}]{Fischer:2011zj}%
  \BibitemOpen
  \bibfield  {author} {\bibinfo {author} {\bibfnamefont {T.}~\bibnamefont
  {Fischer}} \emph {et~al.},\ }\bibfield  {booktitle} {\emph {\bibinfo
  {booktitle} {{Proceedings, 6th International Workshop on Critical Point and
  Onset of Deconfinement (CPOD 2010): Dubna, Russia, August 23-29, 2010}}},\
  }\href {\doibase 10.1134/S1063778812050067} {\bibfield  {journal} {\bibinfo
  {journal} {Phys. Atom. Nucl.}\ }\textbf {\bibinfo {volume} {75}},\ \bibinfo
  {pages} {613} (\bibinfo {year} {2012})}\BibitemShut {NoStop}%
\bibitem [{\citenamefont {Yasutake}\ \emph {et~al.}(2012)\citenamefont
  {Yasutake}, \citenamefont {Noda}, \citenamefont {Sotani}, \citenamefont
  {Maruyama},\ and\ \citenamefont {Tatsumi}}]{Yasutake:2012dw}%
  \BibitemOpen
  \bibfield  {author} {\bibinfo {author} {\bibfnamefont {N.}~\bibnamefont
  {Yasutake}}, \bibinfo {author} {\bibfnamefont {T.}~\bibnamefont {Noda}},
  \bibinfo {author} {\bibfnamefont {H.}~\bibnamefont {Sotani}}, \bibinfo
  {author} {\bibfnamefont {T.}~\bibnamefont {Maruyama}}, \ and\ \bibinfo
  {author} {\bibfnamefont {T.}~\bibnamefont {Tatsumi}},\ }in\ \href
  {https://inspirehep.net/record/1125538/files/arXiv:1208.0427.pdf} {\emph
  {\bibinfo {booktitle} {Recent advances in quarks research}}},\ \bibinfo
  {editor} {edited by\ \bibinfo {editor} {\bibfnamefont {H.}~\bibnamefont
  {Fujikage}}\ and\ \bibinfo {editor} {\bibfnamefont {K.}~\bibnamefont
  {Hyobanshi}}}\ (\bibinfo {year} {2012})\ pp.\ \bibinfo {pages}
  {63--111}\BibitemShut {NoStop}%
\bibitem [{\citenamefont {Most}\ \emph {et~al.}(2018)\citenamefont {Most},
  \citenamefont {Weih}, \citenamefont {Rezzolla},\ and\ \citenamefont
  {Schaffner-Bielich}}]{Most:2018hfd}%
  \BibitemOpen
  \bibfield  {author} {\bibinfo {author} {\bibfnamefont {E.~R.}\ \bibnamefont
  {Most}}, \bibinfo {author} {\bibfnamefont {L.~R.}\ \bibnamefont {Weih}},
  \bibinfo {author} {\bibfnamefont {L.}~\bibnamefont {Rezzolla}}, \ and\
  \bibinfo {author} {\bibfnamefont {J.}~\bibnamefont {Schaffner-Bielich}},\
  }\href@noop {} {\  (\bibinfo {year} {2018})},\ \Eprint
  {http://arxiv.org/abs/1803.00549} {arXiv:1803.00549 [gr-qc]} \BibitemShut
  {NoStop}%
\end{thebibliography}%


%
\end{document}